\documentclass[12pt,preprint]{revtex4}  
\usepackage{amsfonts}
\usepackage{amsmath}
\usepackage{graphicx}
\usepackage{mathcomp}
\usepackage{color}

\newif\iffigures
\figurestrue

\newcommand{\LB}{\linebreak}  

\renewcommand{\vec}{\boldsymbol}

\newcommand{\FINAL}{}

\def\undertilde#1{\mathord{\vtop{\ialign{##\crcr
$\hfil\displaystyle{#1}\hfil$\crcr\noalign{\kern1.5pt\nointerlineskip}
$\hfil\widetilde{}\hfil$\crcr\noalign{\kern1.5pt}}}}}

\usepackage[deletedmarkup=xout]{changes}
\definechangesauthor[color=green]{RPB}
\definechangesauthor[color=cyan]{SW}
\definechangesauthor[color=orange]{JL}
\definechangesauthor[color=blue]{TM}
\definechangesauthor[color=magenta]{MK}

\begin{document}

\title[The mechanism of Hall heating]
{Electron dynamics in planar radio frequency magnetron plasmas:\\ I. The mechanism of Hall heating and the $\boldsymbol{\mu}$-mode}

\author{Denis Eremin$^1$, Dennis Engel$^1$,  Dennis Krüger$^1$, Sebastian Wilczek$^2$, Birk Berger$^2$,  Moritz Oberberg$^2$, Christian Wölfel$^3$, Andrei Smolyakov$^4$, Jan Lunze$^3$,  Peter Awakowicz$^2$, Julian Schulze$^{2}$, Ralf Peter Brinkmann$^1$}

\address{$^1$ Institute of Theoretical Electrical Engineering, Ruhr University Bochum, Germany}
\address{$^2$ Institute of Applied Electromagnetics and Plasma Technology, Ruhr University Bochum, Germany}
\address{$^3$Institute of Automation and Computer Control, Ruhr University Bochum, Germany}
\address{$^4$Department of Physics and Engineering Physics, University of Saskatchewan, Saskatoon, Canada}

\renewcommand{\abstractname}{}

\date{\today}

\begin{abstract}
The electron dynamics and the mechanisms of power absorption in radio-frequency (RF) driven, 
magnetically enhanced capacitively coupled plasmas (MECCPs) at low pressure are investigated. 
The device in focus is a  geometrically asymmetric cylindrical magnetron with a radially nonuniform magnetic field in axial direction and an electric field in radial direction.
The dynamics is studied analytically using the cold plasma model and 
a single-particle formalism, and numerically with the inhouse energy and charge conserving 
particle-in-cell/Monte Carlo collisions code ECCOPIC1S-M. \LB
It is found that the dynamics differs significantly from that of an unmagnetized reference discharge.
\LB
In the magnetized region in front of the powered electrode, 
an enhanced  electric field arises during sheath expansion and
a reversed electric field during sheath collapse. Both fields are needed to ensure  discharge sustaining electron transport against the confining effect of the magnetic field. \LB
The corresponding azimuthal $\mathbf{E}\!\times\! \mathbf{B}$-drift can accelerate electrons into the inelastic energy range which gives rise to a new mechanism of RF power dissipation. It is related to the Hall current and is different in nature from Ohmic heating, 
as which it has been classified in previous literature.
\LB
The new heating is expected to be dominant in many magnetized capacitively coupled discharges.
It is proposed to term it the ``$\mu$-mode'' to separate it from other heating modes.
\end{abstract}

\maketitle

\pagebreak

\section{Introduction} \label{Introduction}

An externally applied magnetic field allows to sustain gas discharges at lower pressures and at higher plasma densities 
than otherwise possible \cite{lieberman_2005}. 
Magnetically enhanced plasmas play a major role in advanced surface processing technologies such as thin film deposition, plasma etching,
or ion implantation \cite{thornton_1978, chapman_1980, LinHinsonClassSandstrom1984,  depla_2008, manos_1989}.
They are often termed ``partially" or ``weakly" magnetized, referring to the fact that -- at typical magnetic flux densities $B$
of up to $100\,\mathrm{mT}$ -- 
\LB only electrons are magnetized, while ions are not.
(A particle is called magnetized when its gyration radius $r_\mathrm{L}$ is smaller than other length scales like the reactor size $L$,
the scale length of the magnetic field $l$, and the mean free path $\lambda$, and its gyration frequency $\omega_\mathrm{c}$ larger that than other relevant frequencies like the modulation 
frequency $\omega_\mathrm{RF}$ or the collision frequency $\nu$ \cite{HagelaarOudini2011}.) \LB
In this study, we focus on radio-frequency (RF) driven magnetrons \cite{thornton_1981,Piel2017},
or, more generally, on magnetically enhanced capacitively coupled plasmas (MECCPs), 
where a radio-frequency voltage drives an electrical current 
across the magnetic field \cite{kushner_2003}. 

In order to optimize and control such discharges, it is crucial to understand 
how exactly  magnetized electrons acquire and utilize their power.
The aspect of ``utilization'' is obvious:  Magnetization can confine energetic electrons to an active region which they can leave only by collisional interaction (classical transport) or by scattering at instability-induced fluctuations of the electric field (anomalous transport). 
This improves utilization of the electron energy.  
In some discharge configurations such as planar magnetrons, the electron drift orbits are closed within the reactor,
 so that the effective system 
size becomes infinite \cite{Rossnagel2020}.  

The processes which enable the electrons to acquire energy are more difficult to address. \LB
In a fluid view, electron heating, as the phenomenon is called,  
requires an electron current~$\vec{j}_\mathrm{e}$\LB in the direction of the electric field $\vec{E}$:
The power dissipation, i.e., the energy flow from the electromagnetic field to the electrons,
 is equal to $P_\mathrm{e} = \vec{j}_\mathrm{e}\!\cdot\!\vec{E}$. 
The mechanisms underlying the transport of electrons in magnetized 
plasmas are diverse and  
notoriously complicated \cite{BoeufSmolyakov2018,Kaganovichetal2020}.\LB
Equally complicated are the mechanisms of electron heating in MECCPs. 
Note that the fluid picture itself may be physically insufficient and should be supplemented by kinetic analysis:
Only those electrons whose kinetic energies are above the inelastic thresholds of chemical reactions can create useful new particles and radiation.
Low-energy electrons do not have this chance and merely take part in transport processes. 
This consideration would motivate to \LB distinguish between electron heating in general and electron energization in particular \cite{anders_2014}.

To get oriented, it helps to first consider the power absorption in non-magnetized CCPs. 
Traditionally, Ohmic heating was assumed: The electrical field 
parallel to the electron flux 
was taken to arise from Ohmic resistance, i.e., from the need to overcome electron inertia and the momentum loss 
connected to elastic collisions
with the neutral background particles \cite{lieberman_2005}. 
This process can already be captured within the Drude model. (We use standard notation:\LB
$m_\mathrm{e}$ is the electron mass, $e$ the electron charge, $n_\mathrm{e}$  the electron density,
$\omega_\mathrm{RF}$ the frequency of the applied voltage, and $\nu_\mathrm{e}= v_\mathrm{the}/\lambda_\mathrm{e}$ the electron neutral collision frequency,
where $\lambda_\mathrm{e}$ is the \LB mean free path and $v_\mathrm{the} = 
\sqrt{8 T_\mathrm{e}/\pi m_\mathrm{e}}$ the thermal speed;
$T_\mathrm{e}$ is the electron temperature.)
In the frequency domain, the plasma conductivity is the complex 
$\underline{\sigma}_\mathrm{e} = e^2 n_\mathrm{e}/(i\omega_\mathrm{RF} + \nu_\mathrm{e})m_\mathrm{e}$. 
%
The electron current is $\underline{\vec{j}}_\mathrm{e} =  \underline{\sigma}_\mathrm{e} \underline{\vec{E}}$, and the 
phase-averaged dissipated power,  
understood to be converted into electron thermal energy, reads:
\begin{equation}
	 \bar{P}_\Omega =  \frac{1}{4}\left( \underline{\vec{j}}_\mathrm{e}\!\cdot\!\underline{\vec{E}}^*+ \underline{\vec{j}}_\mathrm{e}^*\!\cdot\!\underline{\vec{E}}\right)
	=\frac{e^{2}  n_\mathrm{e} \nu_\mathrm{e}}{2 m_\mathrm{e} \left(\nu_\mathrm{e}^2+\omega_\mathrm{RF}^2\right)}\, \underline{\vec{E}}\!\cdot\!\underline{\vec{E}}^*.
\end{equation}   
Kinetic theory gives insight into the underlying mechanism. Assuming that the electric field  is not too strong
-- $e\lambda_\mathrm{e} |\vec{E}| \ll T_\mathrm{e}$, so that the energy increment between collisions is small --, 
Ohmic heating is a diffusion process in energy space. The corresponding diffusion constant is proportional 
to the square of the effective electric field and, provided that the elastic 
collision frequency is smaller than the radio frequency, proportional to the pressure $p$
\cite{tsendin_2011, kudryavtsev_2010}.\LB
At low gas pressure, however, the concept of Ohmic heating loses its explanatory power.\LB
An alternative mechanism, termed ``stochastic heating'',
was proposed by Godyak  \cite{Godyak1972,LiebermanGodyak1998}. 
\LB
The modulated plasma sheath, 
specifically the electron edge, was treated as an oscillating,
specularly reflecting ``hard wall''. As a consequence, if an electron of velocity $v$ 
 collides with the edge
 moving at speed $u_\mathrm{s}$, then the electron velocity after the collision is $2u_\mathrm{s} - v$.  \LB
 Under the  assumption that the oscillation of the electron edge and the trajectories 
 of the individual electrons are not correlated -- which explains the term "stochastic heating" --, \LB this process leads, on~average, to an increase in kinetic energy.
 (It is evident that the stated assumption is problematic, and recent treatments seek to avoid it. However, the name stuck.)\LB
 Further investigations have shown that,  for typical plasma conditions, stochastic heating cannot really be of a collision-free nature. Rather, it must be of a "hybrid" type in that it requires an additional dissipative mechanism in the plasma bulk, for example elastic electron-neutral collisions, even if these are only implicitely accounted for   \cite{Kaganovich1996, Lafleur2015}.

\pagebreak

It was also found that not only the moving sheath edge,
but also the momentarily quasi\-neutral zone behind it contributes to the power dissipation
via the ambipolar field \cite{GozadinosTurnerVender2001,SchulzeDonkoDerzsiKorolovSchuengel2015}. \LB
The concept of ``pressure heating'' is based on the assumption
that sheath heating is caused by the periodic but temporally asymmetric compression and decompression of
the \LB electron fluid near the sheath \cite{GozadinosTurnerVender2001,Turner2009}.
Reference \cite{Lafleur2014} argued that
stochastic heating and pressure heating stem from the same physical mechanism; they differ only in the spatial region where the electron heating is assumed to occur.
For a finite net power absorption, the sheath expansion must not be a ``mirror image'' of the collapse,
so that the dissipation integral $\oint\vec{j}_\mathrm{e}\!\cdot\!\mathrm{d}\vec{E}$ is unequal to zero 
\cite{Turner2009,SchulzeDonkoLafleurWilczekBrinkmann2018}. 

Additional electron heating mechanisms which can be linked to sheath physics are caused by
the action of a reversed electric field during sheath collapse 
(``field reversal heating'') \cite{sato_1990,vender_1992, tochikubo_1992, schulze_2008, eremin_2015} \LB
and the self-excitation of the plasma series resonance (PSR) through sheath-related non\-linearities in asymmetric discharges 
(``nonlinear electron resonance heating'' NERH) \cite{czarnetzki_2006, mussenbrock_2008, lieberman_2008, donko_2009,wilczek_2016,
wilczek_2018}.\LB
Lastly, it was found that also secondary electrons, predominantly of the ion-induced $\gamma$-type, \LB can significantly contribute 
to the electron heating process and even enable a transition of the discharge into 
the $\gamma$-mode \cite{belenguer_1990, godyak_1992}. 
Of course, all listed processes are simultaneously present and may act in synergy \cite{Brinkmann2015}.  

The literature on the subject shows that all heating mechanisms present in unmagnetized 
capacitive plasmas can also be seen in their magnetized counterparts, albeit in modified form.  
An early study by Lieberman, Lichtenberg, and Savas stated that both Ohmic and stochastic heating would be enhanced by a magnetic field,
but via different physical mechanisms \cite{LiebermanLichtenbergSavas1991}.
\LB \color{black}
The increased Ohmic heating was related to a magnetically modified conductivity tensor. 
Stochastic heating would increase because magnetized electrons can  collide multiple times with the sheath during an expansion phase, 
at each collision  picking up additional energy. 
Both heating processes were argued to scale positively with the magnetic field strength,\LB
so that their ratio was predicted to stay constant. 
Another early study found 
increased levels of sheath heating 
under the resonance condition $\omega = 2\,\omega_\mathrm{ce}$ \cite{OkunoOhtsuFujita1994}.
This effect, which requires unusually small magnetic fields, was revisited in \cite{Zhang2021} and \cite{Patil2022}, where it acquired a new name ``electron bounce-cyclotron resonance heating''. Later research came to different results. \LB
Using experiments, particle-in-cell/Monte Carlo collisions simulations, or 
 analytical models,
  references \cite{HutchinsonTurnerDoyleHopkins1995, TurnerHutchinsonDoyleHopkins1996,KrimkeUrbassek1995} claimed that 
even weak magnetic fields could make Ohmic heating prevail over  other heating mechanisms. 
More recent work has argued similarly \cite{BarnatMillerPaterson2008,YouHaiParkKimKimSeongShinLeeParkLeeChang2011,ZhengWangGrotjohnSchuelkeFan2019}.

Also the other mechanisms were seen in magnetized plasmas.
NERH was investigated for different values of the magnetic flux density in RF driven planar magnetron plasmas 
\cite{oberberg_2019, joshi_2018}.
Electric field reversal was also observed and connected with the inhibited electron transport due to the magnetic field
\cite{yeom_1989,krimke_1994,kushner_2003,WangWenHartmannDonkoDerzsiWangSongWangSchulze2020}. 
Influence of ambipolar field heating was seen in \cite{WangWenHartmannDonkoDerzsiWangSongWangSchulze2020}. 
Also secondary electron emission can play a role. 
Kushner concluded that if the magnetic field is oriented parallel to the electrode, 
heating by secondary electrons (of the $\gamma$-type) is most effective when their gyroradius
is comparable to the elastic mean free path \cite{kushner_2003}.\LB
Other research, however, indicated that a presence of secondaries is not always
important for discharge sustainment \cite{zheng_2021}. 

As stated above, all electron heating mechanisms present in unmagnetized capacitive plasmas are also active in MECCPs, albeit in modified form. But does this also apply vice versa?
\LB
Can all electron heating processes in MECCPs be understood 
in terms of their counterparts in unmagnetized CCPs?
We do not believe this is the case.
Instead, we believe that the dominant heating in
MECCPs is "truly magnetic" but   has not been recognized as such. \LB
In particular, we believe that the term ``Hall-enhanced Ohmic heating'' \cite{ZhengWangGrotjohnSchuelkeFan2019}
is a misnomer, 
and that the underlying phenomenon has little in common with the process of Ohmic heating, 
\LB
provided that term is used in the original sense.  
We propose to call it ``Hall heating'' or,  with reference to the role 
of the magnetic field, 
the $\mu$-mode. 
In this study we will out\-line our arguments and support them with analysis and simulation.
Specifically, we will contrast the characteristics of a magnetized discharge with a
nominal field strength of $B=5\,\mathrm{mT}$ with an unmagnetized reference case
of the same geometry. (If necessary, we mark the magnetized and unmagnetized case with (M) and (U), respectively.)

This publication is the first of three companion papers aimed at improving the understanding of the electron dynamics
in magnetized radio frequency discharges operated at low pressures. \LB
It emphasizes the kinetic nature of the electron heating in such discharges 
and demonstrates that the dominant Hall heating is a new mechanism that  
directly leads to efficient ionization. \linebreak
To focus on the electron transport across the magnetic field lines,  
we study a long cylindrical magnetron in axisymmetric approximation.
The second work \cite{eremin_2021} computationally studies a planar RF magnetron in realistic geometry
with electron motion along the magnetic field.

Finally, the third publication \cite{berger_2021} presents experimental observations obtained in a planar RF magnetron and interprets them with help of the newly gained knowledge about the electron dynamics 
in such devices. 


\pagebreak

\section{The device under study and its operation regime \label{s1.5}}

\begin{figure}[h!]\centering
\iffigures	
   \includegraphics[width=\textwidth]{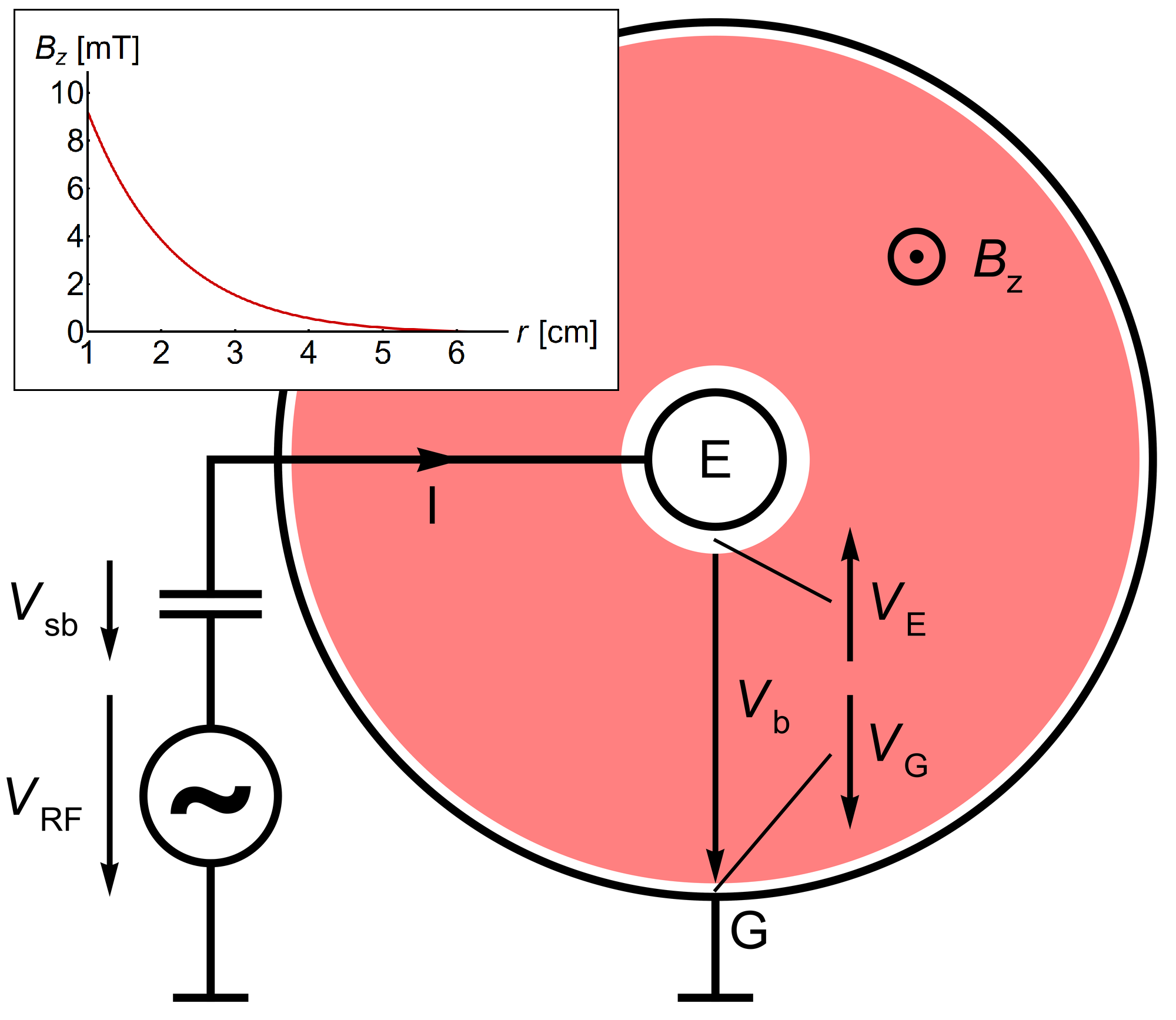} 
\else
\fi
	\caption{The cylindrical magnetron considered in this study. The discharge is driven by an RF voltage source through a blocking capacitor at the inner
electrode, the outer electrode is grounded.\LB The respective radii are $R_\mathrm{E} = 1\,\mathrm{cm}$ and
      $R_\mathrm{G} = 6.5\,\mathrm{cm}$.   Due to the large geometrical asymmetry, \LB
   the plasma sheath at the powered electrode is larger than the 
sheath at the grounded electrode. The magnetic field $B_z(r)$
is axially oriented and is nonuniform in the radial direction (see inset). The arrows indicate the sign conventions for the voltages and currents. \FINAL}
	\label{DischargeSchemeDrawing}
\end{figure}

\begin{figure}[h!]\centering
\iffigures	
   \includegraphics[width=\textwidth]{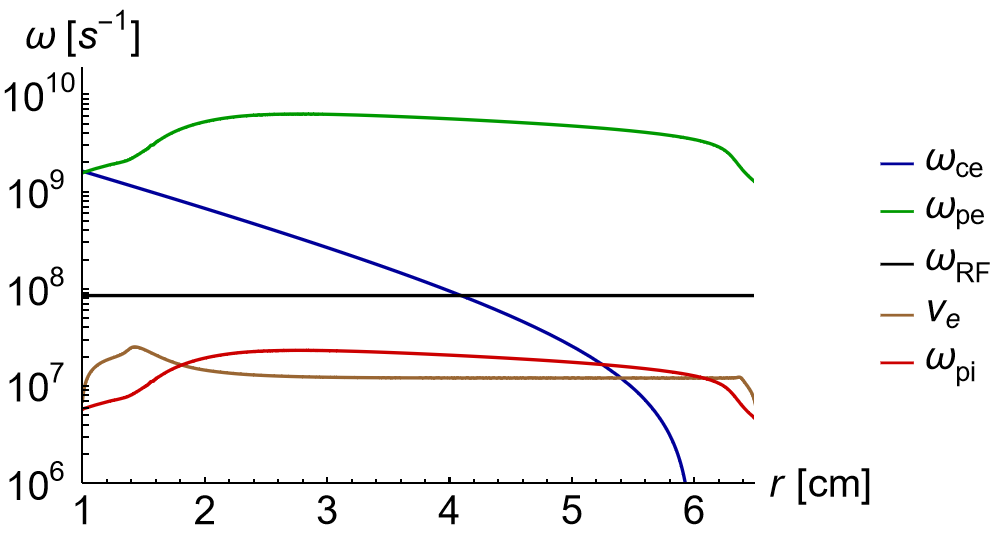}
\else
\fi
	\caption{Characteristic frequencies of the cylindrical magnetron,
	evaluated for the magnetized case.
    Blue: electron gyrofrequency $\omega_\mathrm{ce}$, 
	green: electron plasma frequency $\omega_\mathrm{pe}$,
	black: radio frequency $\omega_\mathrm{RF}$,\LB	
	brown: electron collision frequency $\nu_\mathrm{e}$,
 	red: ion plasma frequency $\omega_\mathrm{pi}$.
	\FINAL}
	\label{FrequencyPlot}
\end{figure}

\begin{figure}[h!]\centering
\iffigures	
   \includegraphics[width=\textwidth]{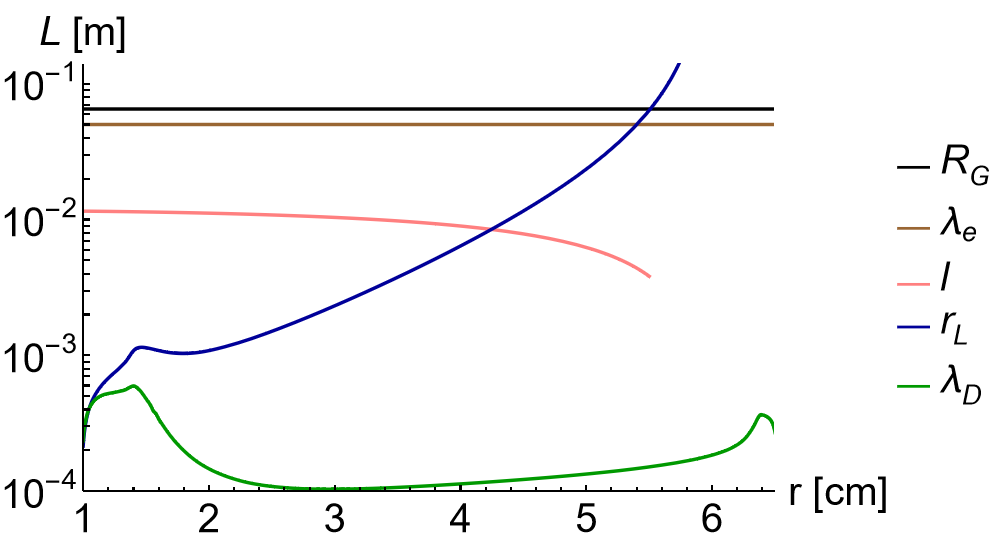} 
\else
\fi
	\caption{Characteristic length scales of the cylindrical magnetron,
	evaluated for the magnetized case. \LB
	Blue: electron gyroradius $r_\mathrm{L}$, black: chamber radius $R_\mathrm{G}$, 
	pink: field gradient scale $l = |B_z/\partial B_z/\partial r|$, \LB
	brown: electron mean free path $\lambda_\mathrm{e}$, 
	green: Debye length $\lambda_\mathrm{D}$.	\FINAL}
	\label{LengthPlot}
\end{figure}

Our research focuses on a  cylindrical magnetron with pronounced geometrical asymmetry.
Fig.~\ref{DischargeSchemeDrawing} provides a schematic and defines the reference directions of the currents and voltages. \LB
The powered inner electrode $\mathrm{E}$ 
has a radius of $R_\mathrm{E} = 1\,\mathrm{cm}$,
the grounded outer electrode $\mathrm{G}$\LB  a radius of $R_\mathrm{G} = 6.5\,\mathrm{cm}$,
with $R_\mathrm{C} = (R_\mathrm{E}+R_\mathrm{G})/2$. 
The electrode area ratio is $A_\mathrm{E}/A_\mathrm{G} = 0.154$.
\LB
Invariance in the axial and azimuthal directions is assumed. 
The first assumption makes the discharge infinitely long. 
(For evaluation purposes, we use a nominal height of $H=10\,\mathrm{cm}$.) 
\LB
The second assumption reflects that symmetry breaking plasma instabilities
are generally not significant for the rather weak magnetic fields ($B \leq 10$ mT) considered in this study \cite{panjan_2019, lucken_2020,Xu_2021}.
\LB
In the radial direction the magnetic field and the plasma characteristics are nonuniform.\LB
The device is operated in argon at a pressure of $p=0.5\,\mathrm{Pa}$
and a temperature of $T=300\,\mathrm{K}$. \LB
A sinusoidal RF voltage with an amplitude of
$\hat{V}=300\,\mathrm{V}$ and a frequency of $f = 13.56\,\mathrm{MHz}$ is 
connected via a large  blocking capacitor,
over which a self bias of $\bar{V}_\mathrm{G} = -161\,\mathrm{V}$ drops.
\LB
The sheath before the powered electrode has a phase-averaged 
thickness of $\bar{s}_\mathrm{E}\approx 0.32\,\mathrm{cm}$;
it develops a large voltage  $\bar{V}_\mathrm{E} = 200\,\mathrm{V}$ that accelerates the impacting ions to high energies. 
\LB
This, in turn, yields a high flux of secondary electrons which are also energized in the sheath.
Analogous processes at electrode $\mathrm{G}$ are less effective,
there $\bar{s}_\mathrm{G} \approx 0.14\,\mathrm{cm}$
and $\bar{V}_\mathrm{G} = 20\,\mathrm{V}$. \LB
The magnetic field $\vec{B} =  B_z(r) \vec{e}_z$ is a 1d model of a planar magnetron field \cite{oberberg_2019,oberberg_2018,oberberg_2020,berger_2021}.
\LB
(Clearly, the model field is not curl-free. 
However, the validity of our analysis is not affected.)
\LB 
Following the logic of reference \cite{trieschmann_2013}, we adopt the 
axial profile of the radial magnetic field\LB component 
directly above the racetrack  from \cite{eremin_2021}, see Fig.~\ref{DischargeSchemeDrawing}.
An analytical form is as follows,
where $B_i = \{-1.46,21.98,-0.90,9.92\}\,{\rm mT}$
and $\lambda_i = \{35.4,81.2,127.3,173.4\}\,\mathrm{m}^{-1}$:
\begin{align} 
    B_z(r) = \sum\limits_{i=1}^4 B_i \exp(-\lambda_i r). \label{MagneticField}
\end{align}
Figs.~\ref{FrequencyPlot} and \ref{LengthPlot} show the
frequency and length scales of the discharge.
In the magnetized region, \LB
the electron gyro frequency and the plasma frequency are  
comparable, $\omega_\mathrm{ce} \approx \omega_\mathrm{pe}  \approx 10^9\,\mathrm{s}^{-1}$,
while the radio frequency and the collision rate are smaller, 
$\omega_\mathrm{RF} \approx 10^8\,\mathrm{s}^{-1}$
and $\nu_\mathrm{e} \approx 10^7\,\mathrm{s}^{-1}$. 
Likewise, the gyroradius and the Debye length are the smallest scales, 
$r_\mathrm{L} \approx \lambda_\mathrm{D} \approx 5\!\times\! 10^{-4}\,\mathrm{m}$,
the gradient scale is larger, $l \approx 5\!\times\! 10^{-3}\,\mathrm{m}$;
and the mean free path even more, $\lambda_\mathrm{e} \approx 5\!\times\! 10^{-2}\, \mathrm{m}$. 
The electron thermal speed and the $E\!\times\! B$ 
drift speed are similar,
$v_\mathrm{the} \approx v_{E\times B}
\approx 10^6 \, \mathrm{m/s}$.
The ions are not affected by the magnetic field.
(Data from the simulations below.)

\pagebreak

\section{Analysis based on the cold plasma model \label{cpm}}

A first insight into the discharge dynamics is obtained by considering the cold plasma model, also known as the Drude model \cite{lieberman_2005}.
We adopt the cylindrical geometry as discussed above.
The electron plasma frequency $\omega_\mathrm{pe}(r)$ and the electron gyro frequency 
$\omega_\mathrm{ce}(r)$ depend on $r$;\LB
the elastic electron collision frequency $\nu_\mathrm{e}$ is constant. 
The equations of continuity and motion for the charge density $\rho(r,t)$ and for 
the non-vanishing components $j_r(r,t)$ and $j_\theta(r,t)$ of the electric current density 
are
\begin{align}
    &\frac{\partial\rho}{\partial t} + 
     \frac{1}{r} \frac{\partial}{\partial r} \left(r j_r\right) = 0,\\[0.5ex]
    & \frac{\partial j_r}{\partial t}  = \varepsilon_0 \omega_\mathrm{pe}^2 E_r -\omega_\mathrm{ce} j_\theta  - \nu_\mathrm{e} j_r, \\[0.5ex]
&\frac{\partial j_\theta}{\partial t}  = \omega_\mathrm{ce} j_r -\nu_\mathrm{e} j_\theta.    
\end{align}
The electric field $\vec{E}=E_r(r,t)\vec{e}_r$, derived from a potential $\Phi(r,t)$, 
follows Poisson's equation
\begin{align}
    \varepsilon_0   \frac{1}{r} \frac{\partial}{\partial r} \left(r E_r\right)
      = - \varepsilon_0   
      \frac{1}{r} \frac{\partial}{\partial r} \left(r \frac{\partial\Phi}{\partial r}\right)
       =  \rho.
\end{align}
As a consequence of the equations, the sum of the particle current and the 
displacement current is divergence-free. This allows to define the 
current through the discharge as
\begin{align}
    I(t) = 2\pi r H \left( j_r + \varepsilon_0 \frac{\partial E_r}{\partial t}\right). 
    \label{CurrentField}
\end{align}
If the time-evolution is harmonic with frequency $\omega$
(which is in this context not necessarily the applied radio frequency
$\omega_\mathrm{RF}$), 
the complex currents $\underline{j}_r$ and $\underline{j}_\theta$ can be expressed as follows.\LB
The expressions after the tilde hold in the limiting case 
         $\nu = 0$, $\omega \ll \omega_\mathrm{ce}$:
\begin{align}
&\underline{j}_r = 
   -\frac{(i\omega +\nu_\mathrm{e})\varepsilon_0\omega_\mathrm{pe}^2  }
   {\omega^2-\omega_\mathrm{ce}^2 -\nu_\mathrm{e}^2  - 2 i \nu_\mathrm{e}  \omega}  \underline{E}_r
   \sim  \frac{i\omega\varepsilon_0\omega_\mathrm{pe}^2  }
   {\omega_\mathrm{ce}^2 }  \underline{E}_r = e n_\mathrm{e}  \frac{i\omega \underline{E}_r}{\omega_\mathrm{ce} B_z}, \label{jr}\\[0.5ex]
&\underline{j}_\theta =   - \frac{\omega_\mathrm{ce} \varepsilon_0\omega_\mathrm{pe}^2}{\omega^2-\omega_\mathrm{ce}^2 -\nu_\mathrm{e}^2  - 2 i \nu_\mathrm{e}  \omega}  \underline{E}_r\sim 
   \frac{\varepsilon_0\omega_\mathrm{pe}^2}{\omega_\mathrm{ce}}  \underline{E}_r
   =  e n_\mathrm{e}  \frac{\underline{E}_r}{B_z}  . \label{jtheta}
\end{align}
The relation of the discharge current to the electric field can be written as
\begin{align}
    \underline{I} = 2 \pi r  H \, i \omega \varepsilon_0 \varepsilon_\mathrm{p} \underline{E}_r,
\end{align}
where the relative dielectric constant of the plasma is
\begin{align}
     \varepsilon_\mathrm{p} = 
     1-\frac{\omega_\mathrm{pe}^2 (\omega -i \nu_\mathrm{e} )}{\omega 
   \left(\omega^2-\omega_\mathrm{ce}^2-\nu_\mathrm{e}^2 - 2 i \nu_\mathrm{e}  \omega\right)}
   \approx
   -\frac{\omega_\mathrm{pe}^2 (\omega -i \nu_\mathrm{e})}{\omega 
   \left(\omega^2-\omega_\mathrm{ce}^2-\nu_\mathrm{e}^2 - 2 i \nu_\mathrm{e}  \omega\right)}.
\end{align}

The local resonance condition $\varepsilon_\mathrm{p}= 0$ (with full form) leads 
to a current-free resonance at the upper hybrid frequency
$\omega_\mathrm{uh} = \sqrt{\omega_\mathrm{pe}^2+\omega_\mathrm{ce}^2}$,
often termed "plasma parallel resonance" \cite{wilczek_2018}. \LB
If the approximate form of $\varepsilon_\mathrm{p}$ is used, this fast local phenomenon is excluded and only the slower global behavior remains in the description.
Solving \eqref{CurrentField} for the electric field in terms of the 
global current $I$ and integrating over the bulk -- from $R_\mathrm{E}+ \bar{s}_\mathrm{E}$
to $R_\mathrm{G}- \bar{s}_\mathrm{G}$ -- yields the voltage drop over the bulk. 
Added to this is the capacitive voltage drop over the sheaths. \LB
We evaluate the model for the unmagnetized reference case (U) and the magnetized case (M)\LB described 
in sections \ref{SectionWithoutMagneticField} and \ref{SectionWithMagneticField}
below.
For a compact notation, we introduce the inverse capacitances of the two sheaths and the inertia coefficient ("inductance") of the bulk:
\begin{align}
&    C_\mathrm{E}^{-1} = \int_{R_\mathrm{E}}^{R_\mathrm{E}+ \bar{s}_\mathrm{E}} \frac{1}{2\pi r H \varepsilon_0} \,\mathrm{d}r \;\; =  1.32\times 10^{11}\,\mathrm{V/As}\, \mathrm{(U)} \;\; \mathrm{or}\;\;
 5.00\times 10^{10}\,\mathrm{V/As}\, \mathrm{(M)}, \\[0.5ex] 
&     C_\mathrm{G}^{-1} =   \int_{R_\mathrm{G}-\bar{s}_\mathrm{G}}^{R_\mathrm{G}} \frac{1}{2\pi r H \varepsilon_0} \,\mathrm{d}r \;\; =  1.63\times 10^{10}\,\mathrm{V/As}\, \mathrm{(U)} \;\; \mathrm{or}\;\;
 3.90\times 10^{9}\,\mathrm{V/As}\, \mathrm{(M)},\\[0.5ex]
  &  L_\mathrm{B} = \int_{R_\mathrm{E}+ \bar{s}_\mathrm{E}}^{R_\mathrm{G}+ \bar{s}_\mathrm{G}}
     \frac{1}{2 \pi  r H \varepsilon_0 \omega_\mathrm{pe}^2(r)} \, \mathrm{d}r \;\;=  1.01\times 10^{-7}\,\mathrm{V/As}\, \mathrm{(U)} \;\; \mathrm{or}\;\;
 1.66\times 10^{-8}\,\mathrm{Vs/A}\, \mathrm{(M)}. 
\end{align}
In the magnetized case there is an additional coefficient, 
formally an inverse "capacitance".

Physically, this capacitive behavior can be understood as follows: The magnetized electrons in the bulk experience  $\vec{E}\!\times\!\vec{B}$ drift, so that their kinetic energy density is 
$m_\mathrm{e} \left({E_r}/{B_z}\right)^2   n_\mathrm{e}/2$.
The energy density of an electric field is
$\varepsilon_0\varepsilon_\mathrm{r} E_r^2/2$. 
Formally equating the two yields a 
"relative permittivity" of $\varepsilon_\mathrm{r}=\omega_\mathrm{pe}^2/\omega_\mathrm{ce}^2$.
Applying the formula to the series connection of an infinite number of 
infinitely thin
capacitors leads to
\begin{align}
    C_\mathrm{B}^{-1} = \int_{R_\mathrm{E}+ \bar{s}_\mathrm{E}}^{R_\mathrm{G}+ \bar{s}_\mathrm{G}} 
     \frac{  \omega_\mathrm{ce}^2(r)}{2 \pi  r H \varepsilon_0 \omega_\mathrm{pe}^2(r)}
     \, \mathrm{d}r \;\; =  0\,\mathrm{V/As}\, \mathrm{(U)} \;\; \mathrm{or}\;\;
 9.64\times 10^{9}\,\mathrm{V/As}\, \mathrm{(M)}. \label{CGyro}
\end{align}

The impedance of the discharge is then
\begin{align}
     Z(\omega) = \frac{C_\mathrm{E}^{-1}}{i\omega }   +    \frac{C_\mathrm{G}^{-1}}{i\omega } 
     +     (i\omega + \nu_\mathrm{e}) L_\mathrm{B}
     +\frac{C_\mathrm{B}^{-1}}{(i\omega + \nu_\mathrm{e})}.   
\end{align}
At the frequency $\omega_\mathrm{RF}=2\pi\times 13.56\,\mathrm{MHz}$, 
the overall impedance is dominated by the sheaths\LB 
and acts as a lossy capacitor,
$Z(\omega_\mathrm{RF})=  \left(1.22 - 1726\,\mathrm{i}\right)\,\mathrm{V/A} \, 
\mathrm{(U)} \; \mathrm{or} \;
\left(15.9 - 742\,\mathrm{i}\right)\,\mathrm{V/A}$\, (M). \LB
This implies that the system response is nearly quasistatic:
The discharge state is controlled  
by the momentary value $V_\mathrm{RF}(t)$, and the discharge current reflects the derivative $\mathrm{d}V_\mathrm{RF}/\mathrm{d}t$. 
\LB
Superimposed on this, however, can be an excitation of the plasma series resonance (PSR).  \LB Its frequency is obtained by setting the impedance $Z(\omega)$ equal to zero and solving for $\omega$:
\begin{align}
    \omega_\mathrm{PSR} = \sqrt{\left(C_\mathrm{E}^{-1}+C_\mathrm{G}^{-1}+C_\mathrm{B}^{-1}\right)/
                 {L_\mathrm{B}}}
    \;\; =  1.21\times 10^9\,\mathrm{s}^{-1}\, \mathrm{(U)} \;\;  \mathrm{or} \;\;
 1.96\times 10^{9}\,\mathrm{s}^{-1}\,  \mathrm{(M)}. \label{PSR}
\end{align}
In the unmagnetized case, the ratio of the PSR frequency to the radio frequency is about 14;
\LB
in the magnetized case, the ratio is roughly 23. This shift can be understood in 
elementary terms by assuming also uniform plasma density and uniform magnetic flux density profiles.\LB Eq.~\eqref{PSR} then states  $\omega_\mathrm{PSR} \mathrm{(M)} = \sqrt{\omega_\mathrm{PSR}^2\mathrm{(U)} + \omega_\mathrm{ce}^2}$, 
where $\omega_\mathrm{PSR}\mathrm{(U)} = \sqrt{(\bar{s}_\mathrm{E}+\bar{s}_\mathrm{G})/d}\;\omega_\mathrm{pe}$ \cite{Qiu2001}.\LB
The resonances are weakly damped
with $\beta = 6.0\times 10^{6}\,\mathrm{s}^{-1}\, \mathrm{(U)}
\;\; \mathrm{or}\;\; 7.0\times 10^{6}\,\mathrm{s}^{-1}\, \mathrm{(M)}$.

\pagebreak

\section{Single-particle picture \label{sp}}

An alternative picture can be established by analyzing the motion of individual electrons.
We assume a  magnetic field $\vec{B}= B_z(r)\,\vec{e}_z$ 
and an RF modulated electric field 
$\vec{E} = E_r(r,t)\,\vec{e}_r$. \LB
For our numerical trajectory example, we take the magnetic field \eqref{MagneticField} and the electrical field from the simulations below,
see Fig.~\ref{ElectricFieldMagneticCase}.
(The results of this section can thus be compared directly with the simulation outcome.)
The motion along the magnetic field lines is simple: 
The coordinate $z$ moves uniformly in time, the velocity $v_z$ is constant:
\begin{align}
& \frac{\mathrm{d}z}{\mathrm{d}t} = v_z,   \label{eq3_1}   \\[0.5ex] 
&\frac{\mathrm{d}v_z}{\mathrm{d}t} =0. 
\end{align}
Consequently, also the kinetic energy of the 
motion along the magnetic field is constant:
\begin{align}
  \frac{\mathrm{d}\epsilon_\parallel}{\mathrm{d}t} 
  \equiv \frac{\mathrm{d}\epsilon_z}{\mathrm{d}t}
    \equiv \frac{\mathrm{d}}{\mathrm{d}t}
     \left(\frac{1}{2}m_\mathrm{e}v_z^2\right)
     = 0.
\end{align}
The motion in the 
$r$-$\theta$-plane
is governed by the Lorentz force and the inertial pseudo-forces,
with the former typically dominating the latter by an order of
magnitude:
\begin{align}
& \frac{\mathrm{d}r}{\mathrm{d}t} = v_r,   \label{eqr}   \\[0.5ex] 
&\frac{\mathrm{d}\theta}{\mathrm{d}t} = \frac{v_\theta}{r},  \label{eqtheta}
\\[0.5ex] 
&\frac{\mathrm{d}v_r}{\mathrm{d}t}=-\frac{e}{m_\mathrm{e}}(E_r+v_\theta B_z)+\frac{v_\theta^2}{r}, \label{eqvr}
\\[0.5ex] 
 &\frac{\mathrm{d}v_\theta}{\mathrm{d}t} = \frac{e}{m_\mathrm{e}} v_r B_z - \frac{v_r v_\theta}{r}.  \label{eqvtheta}
\end{align}
The kinetic energy in the $r$-$\theta$-plane
is modulated by the power influx from the electric field.
As expected, the Lorentz force and the inertial force do not contribute:
\begin{align}
\frac{\mathrm{d}\epsilon_\perp}{\mathrm{d}t}\equiv
  \frac{\mathrm{d}}{\mathrm{d}t}\left(\epsilon_r+\epsilon_\theta\right)
    \equiv \frac{\mathrm{d}}{\mathrm{d}t}
     \left(\frac{1}{2}m_\mathrm{e}v_r^2+\frac{1}{2}m_\mathrm{e}v_\theta^2\right)
     = - e v_r E_r. \label{rEnergylaw}
\end{align}

\begin{figure}[h!]\centering
\iffigures	
   \includegraphics[width=\textwidth]{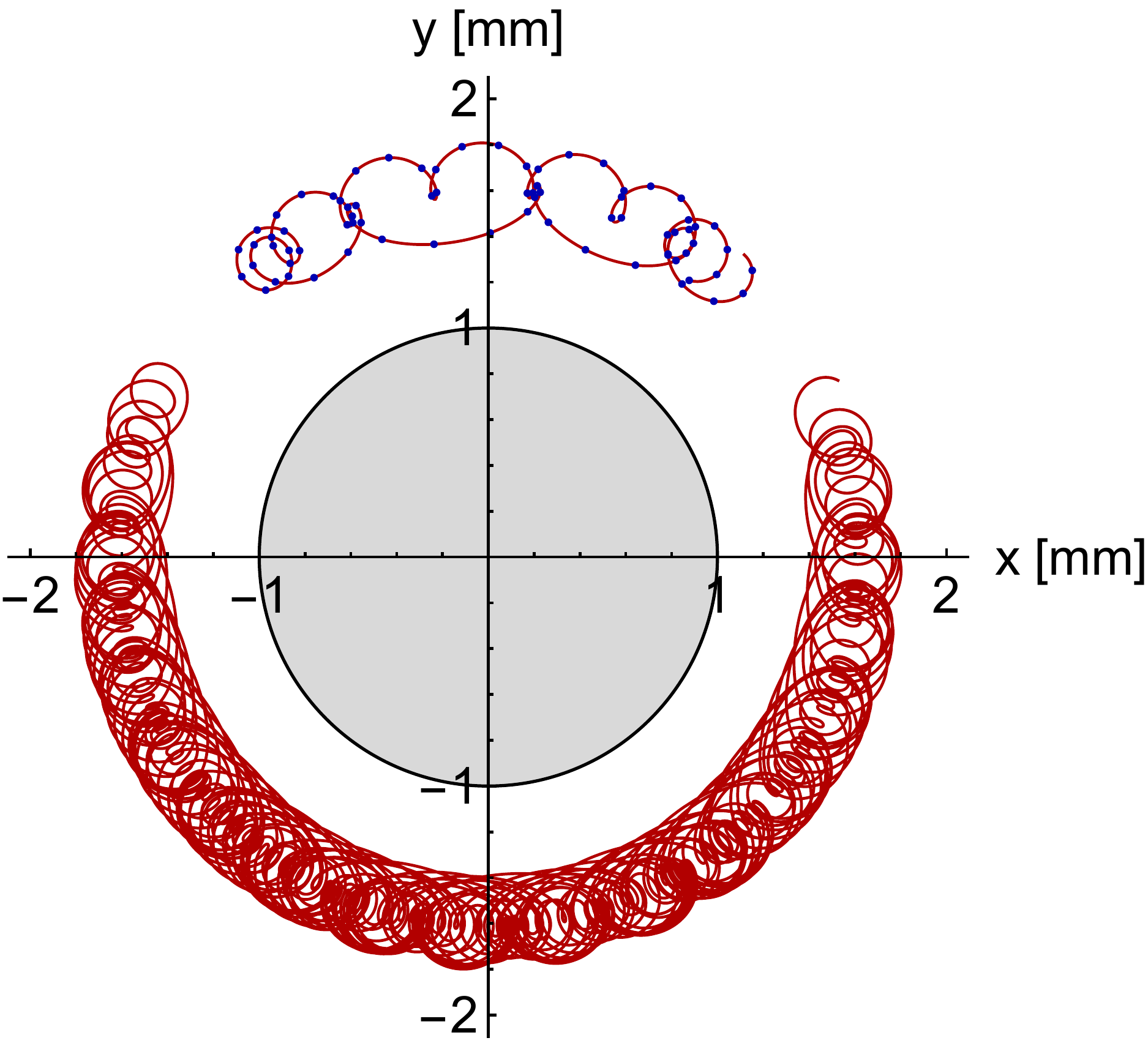}
\else
\fi
	\caption{Trajectory of a magnetized electron in the cross-sectional plane, showing the superposition of small scale gyration and large scale drift. 
	Top part: Trajectory recorded over one RF period; the blue dots have $1\,\mathrm{ns}$
	separation. Bottom part: Trajectory followed over 25 RF periods. \FINAL}
	\label{TrajectoryPlot}
\end{figure}

\begin{figure}[h!]\centering
\iffigures	
   \includegraphics[width=15cm]{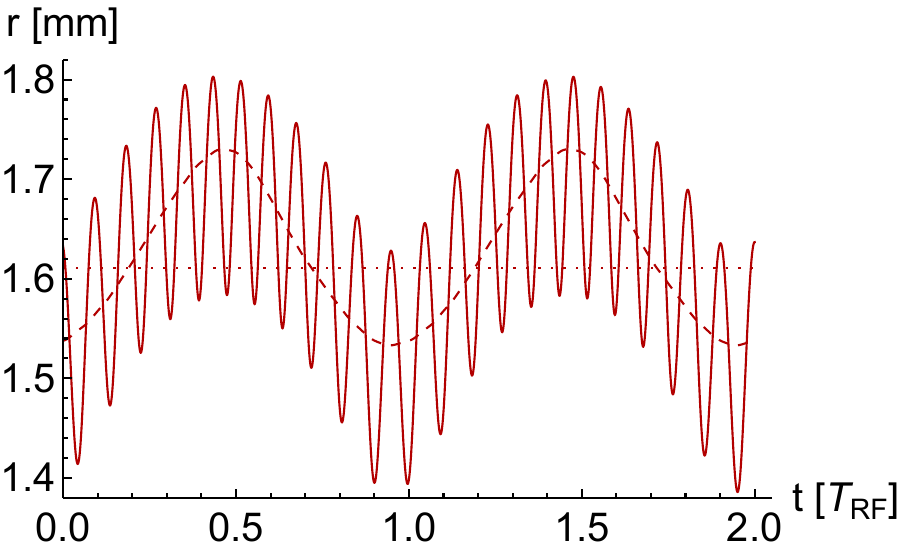} 
\else
\fi
\caption{Solid red: Radial coordinate $r(t)$ of the sample
         trajectory followed over two RF periods.
         Dotted red: Reference radius $\hat{r}$.
	     Dashed red: Position of the guiding center
	     $\hat{r} - \hat{v}_{E\times B}(t)/\bar{\Omega}$. \FINAL}
	\label{rTrajectory}
\end{figure}

\begin{figure}[h!]\centering
\iffigures	
  \includegraphics[width=\textwidth]{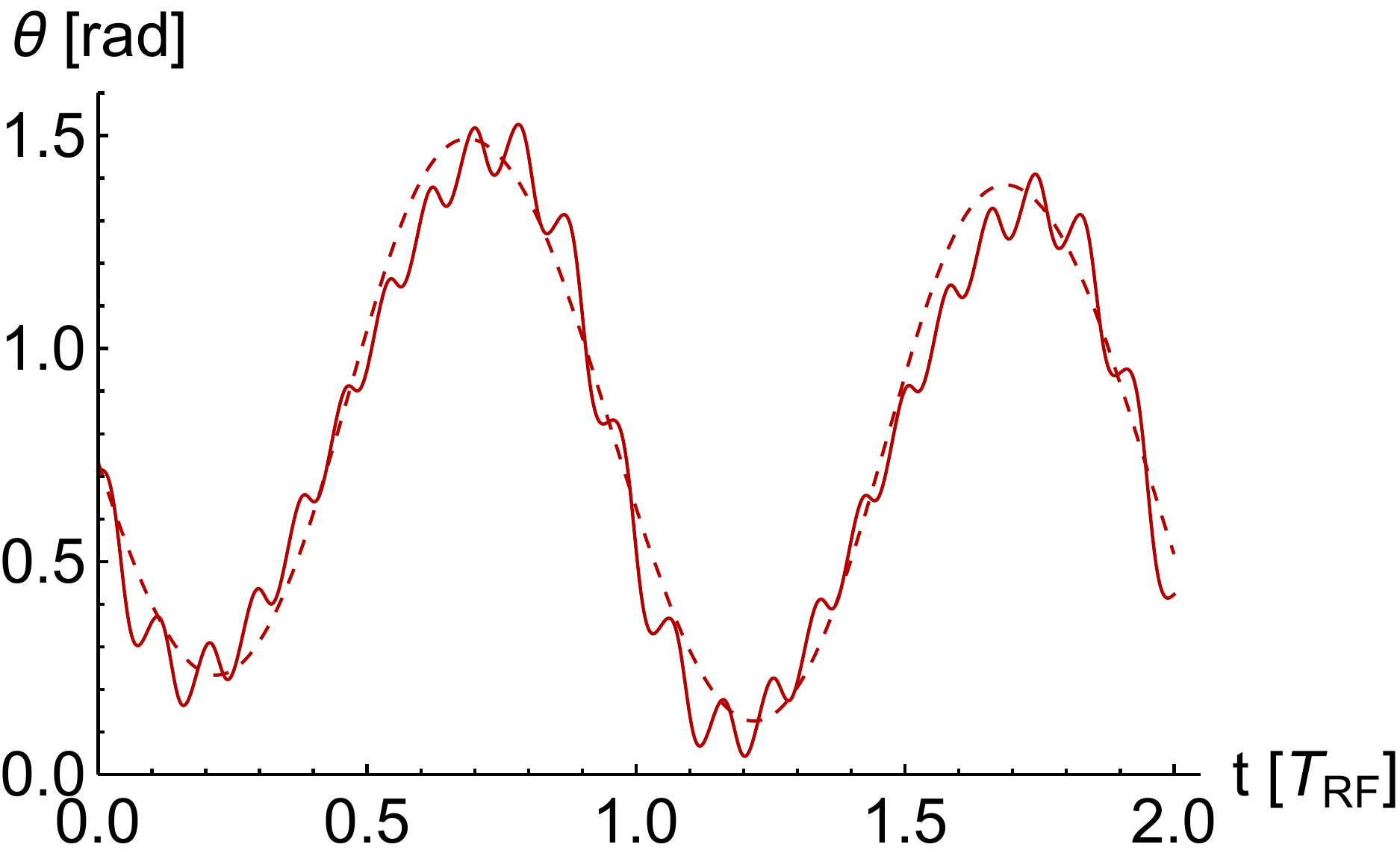}
\else
\fi
	\caption{Solid red: Azimuthal coordinate $\theta(t)$ of the sample
         trajectory followed over two RF periods.
         Dashed red: Position of the guiding center
	     $\hat\theta$. \FINAL}
	\label{thetaTrajectory}
\end{figure}

\begin{figure}[h!]\centering
\iffigures	
   \includegraphics[width=\textwidth]{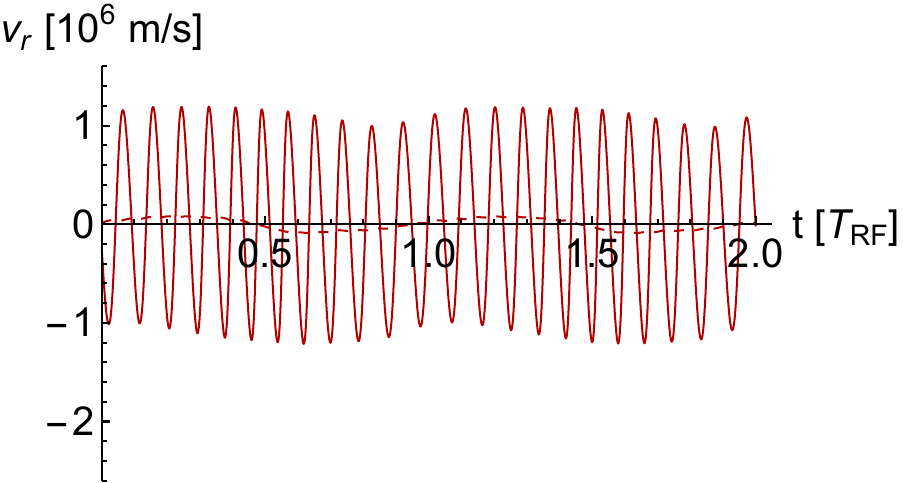} 
\else
\fi
	\caption{Solid red: Radial velocity $v_r$ of the sample trajectory 
	          recorded over two RF periods.
	Dashed red: Drift velocity due to polarization drift. \FINAL}
	\label{vrTrajectory}
\end{figure}

\begin{figure}[h!]\centering
\iffigures	
   \includegraphics[width=\textwidth]{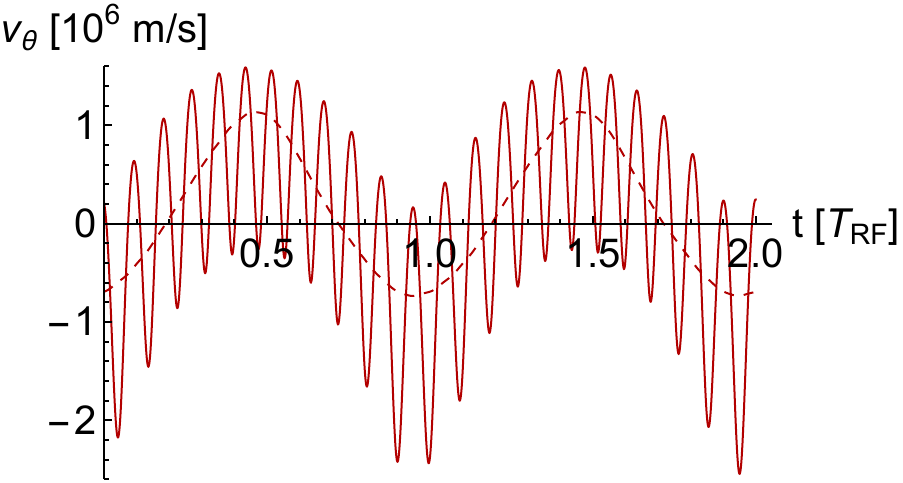} 
\else
\fi
	\caption{Solid red: Azimuthal Velocity $v_\theta$ of the sample trajectory
	recorded over two RF periods.
	Dashed red:  Drift velocity due to $\vec{E}\!\times\!\vec{B}$ drift.
\FINAL}
	\label{vthetaTrajectory}
\end{figure}

\begin{figure}[h!]\centering
\iffigures	
   \includegraphics[width=\textwidth]{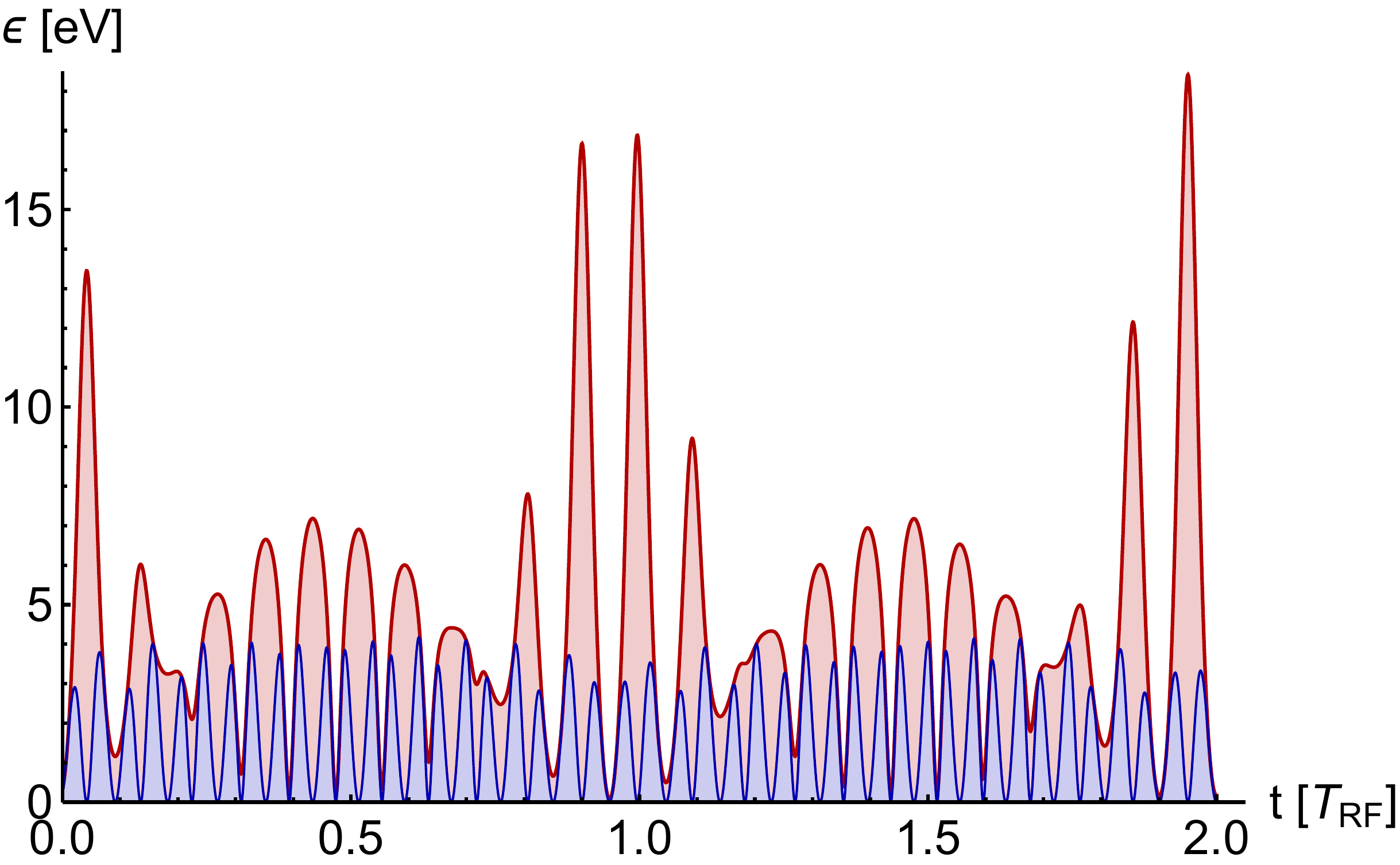}
\else
\fi
	\caption{Kinetic particle energy in radial direction (blue) and in azimuthal direction (red) of the sample trajectory over 
	two RF periods. The second curve shows constructive and destructive interference of gyro motion and drift. \FINAL}
	\label{EnergyPlot}
\end{figure}

A numerical integration of the cross-sectional 
equations of motion 
can easily be carried out. For a medium-energy particle in the strongly magnetized zone,
Fig.~\ref{TrajectoryPlot} diplays the trajectory in the cross-sectional $r$-$\theta$-plane
over a single RF period and over a stretch of 25 RF periods. \LB
It clearly shows a superposition of gyromotion and drift.
Figs.~\ref{rTrajectory} and \ref{thetaTrajectory} display the evolution of the 
coordinates $r(t)$ and $\theta(t)$ over two RF periods, respectively, while Figs.~\ref{vrTrajectory} and \ref{vthetaTrajectory} show the velocities $v_r(t)$ and $v_\theta(t)$. Fig.~\ref{EnergyPlot} shows the corresponding kinetic energies.  

It is instructive to compare the numerically calculated trajectories above with approximate solutions constructed by means of perturbation analysis. 
We apply a technique based on a series expansion in  the ratio of the gyroradius to the discharge scale, $\eta = r_\mathrm{L}/|r-R_\mathrm{E}| \approx 0.1$.
For a qualitative picture, it suffices to evaluate the power series up to the leading order in $\eta$. \LB
(Appendix A provides more mathematical details and formulates also higher order terms.) \LB
We start by noting that there is,
 due to the cylindrical symmetry,  
a strict constant of motion, \LB
the canonical momentum $p_\theta =  r m_\mathrm{e} v_\theta  - r e A_\theta(r) $,
where $A_\theta$ is the $\theta$-component of the 
magnetic vector potential $\mathbf{A}$. 
It can be used to define
the temporally constant reference radius $\hat{r}$ of an electron trajectory via the relation
$p_\theta  \stackrel{!} =  - \hat{r}\, e A_\theta(\hat{r})$.
In the numerical example, $\hat{r} = 16.1\,\mathrm{mm}$.\LB
We use the circumflex also to denote the constant magnetic flux density  $\hat{B}= B_z(\hat{r})=5.4\,\mathrm{mT}$ 
at the reference radius  $\hat{r}$.
The corresponding gyrofrequency is $	\hat{\Omega} =  e \hat{B}/m_\mathrm{e}=9.5\times 10^8\,\mathrm{s}^{-1}$.\LB
The temporally varying electric field strength at the reference point is
termed $\hat{E}(t) = E_r(\hat{r},t)$;\LB
the corresponding drift speed is
$\hat{v}_{E\times B}(t)=\hat{E}(t)/\hat{B}$.
An approximate solution of the equations of motion can then be
expressed as follows. Note the correspondence of  \eqref{vrMotion}
and \eqref{vthetaMotion} with \eqref{jr} and \eqref{jtheta} under
the assumptions $\nu_\mathrm{e}\to 0$, $\omega\ll \omega_\mathrm{ce}$, and 
$\hat{\rho}\to 0$: 
\begin{align}
    & r = \hat{r}   -\frac{\hat{v}_{E\times B}}{\hat{\Omega}}+ \hat{\rho}\cos\bigl(\hat{\Omega}\,t +\hat{\phi}\bigr), \label{rMotion}\\
		& \theta = \hat{\theta}  +  
		\frac{\hat{\rho}}{\hat{r}} \sin\bigl(\hat{\Omega}\,t +\hat{\phi}\bigr), \label{thetaMotion}\\
	  &v_r =  -\frac{1}{\hat{\Omega}} \frac{\mathrm{d}\hat{v}_{E\times B}}{\mathrm{d}t}- 
		      \hat{\Omega}\,\hat{\rho}\, \sin\bigl(\hat{\Omega}\,t +\hat{\phi}\bigr), \label{vrMotion}\\
		&v_\theta= -\hat{v}_{E\times B}(t)+\hat{\Omega} \,\hat{\rho} \, \cos\bigl(\hat{\Omega}\,t +\hat{\phi}\bigr). \label{vthetaMotion}
\end{align}
The gyroradius $\hat{\rho}$ is an adiabatic constant, i.e., independent
of time, while the gyrophase offset $\hat\phi$ and the azimuth offset $\hat{\theta}$ exhibit a slow evolution. In the next order approximation,
this is described as follows, with the prime indicating 
the derivative with respect to $\hat{r}$: 
\begin{align}
& \frac{\mathrm{d}\hat{\phi}}{\mathrm{d}t} = 
\frac{3}{2} 
\left(\frac{1}{\hat{r}}  -\frac{\hat{B}^\prime}{\hat{B}}  \right) \hat{v}_{E\times B} +\frac{\hat{E}^\prime}{2 B}, \\[0.5ex]
& \frac{\mathrm{d}\hat{\theta}}{\mathrm{d}t}= 
    -\frac{1}{\hat{r}}\hat{v}_{E\times B}.
\end{align}
The kinetic energy of the electron can be calculated as
\begin{align}
\epsilon  \approx 
\frac{1}{2}&m_\mathrm{e}\hat{\Omega}^2\hat{\rho}^2 
+\frac{1}{2}m_\mathrm{e}\left( \hat{v}_{E\times B}(t)^2+\frac{1}{\hat{\Omega}^{2}}\left(\frac{\mathrm{d}\hat{v}_{E\times B}}{\mathrm{d}t}\right)^2\right) \label{KineticEnergy}
\\[0.5ex]\nonumber
 &+m_\mathrm{e}\hat{\Omega} \hat{\rho}\left(-\cos\bigl(\hat{\Omega}\,t +\hat{\phi}\bigr) \hat{v}_{E\times B}(t)+\sin\bigl(\hat{\Omega}\,t +\hat{\phi}\bigr) 
\frac{1}{\hat{\Omega}}\frac{\mathrm{d}\hat{v}_{E\times B}}{\mathrm{d}t}
\right)+ \frac{1}{2} m_\mathrm{e} v_z^2.
\end{align}

Obviously, the electron dynamics can be described as the superposition of a large-scale drift of the guiding center -- 
first terms in \eqref{rMotion} - \eqref{vthetaMotion} --
and a small-scale gyromotion (last terms). 
The guiding center trajectories have a banana-like shape, 
which can be explained physically:
\LB
The force $\vec{F} = - e\hat{E}\vec{e}_r$ perpendicular to $\vec{B}$ modulates the energy of the gyrating electron;\LB
it grows when the electron moves in the direction of $\vec{F}$ 
and decreases when it moves otherwise. \LB
The net effect is a drift $\vec{v}_\mathrm{d} = -\hat{v}_{E\times B}(t) \vec{e}_\theta$. The resulting shift in the azimuthal position $r\Delta\theta$ is of the order of $v_{E\times B}/\omega_\mathrm{RF}$.
In contrast, the motion in the radial direction is of second order: 
\LB
Because the drift $\vec{v}_\mathrm{d}$ is RF modulated, 
it causes an acceleration that acts as a pseudo-force.\LB
The corresponding drift is known as the polarization drift
\cite{lieberman_2005},
it has a phase shift of 90° and is an order of magnitude smaller 
than  the $E\!\times\! B$ drift. The corresponding shift $	\Delta\vec{r}_\mathrm{p}$
in the radial position is proportional to the electric field:
\begin{align}
	&\vec{v}_\mathrm{p} = 
	-m_e \frac{\mathrm{d}\hat{v}_{E\times B} }{\mathrm{d}t}\vec{e}_\theta \!\times\! \frac{\hat{\vec{B}}}{e\hat{B}^2} =
	 -\frac{1}{\hat{\Omega}} \frac{\mathrm{d}\hat{v}_{E\times B}}{\mathrm{d}t}\, \vec{e}_r, \\[0.5ex]
&	\Delta\vec{r}_\mathrm{p} = 
	 -\frac{\hat{v}_{E\times B}}{\hat{\Omega}} \, \vec{e}_r =
	 - \frac{m_\mathrm{e}\hat{E}}{e \hat{B}^2} \, \vec{e}_r.
	 \label{Deltarp}
\end{align}
The ratio $\Delta r_\mathrm{p}/r\Delta\theta$ thus scales as
$\omega_\mathrm{RF}/\omega_\mathrm{ce} \ll 1$.
The banana shape results from the curvature \LB of the azimuthal coordinate lines. The trajectories somewhat resemble the banana-shaped trajectories of the guiding center of trapped particles in tokamaks, although the
underlying drift mechanisms are of a different nature \cite{wesson_2011}.
Note that the azimuthal drift velocity and the gyro speed are comparable.  
The azimuthal kinetic energy $\epsilon_r$ therefore shows destructive and constructive interference; values of nearly $20\,\mathrm{eV}$ can be reached,
see Fig.~\ref{EnergyPlot}.

\begin{figure}[h!]\centering
\iffigures	
   \includegraphics[width=\textwidth]{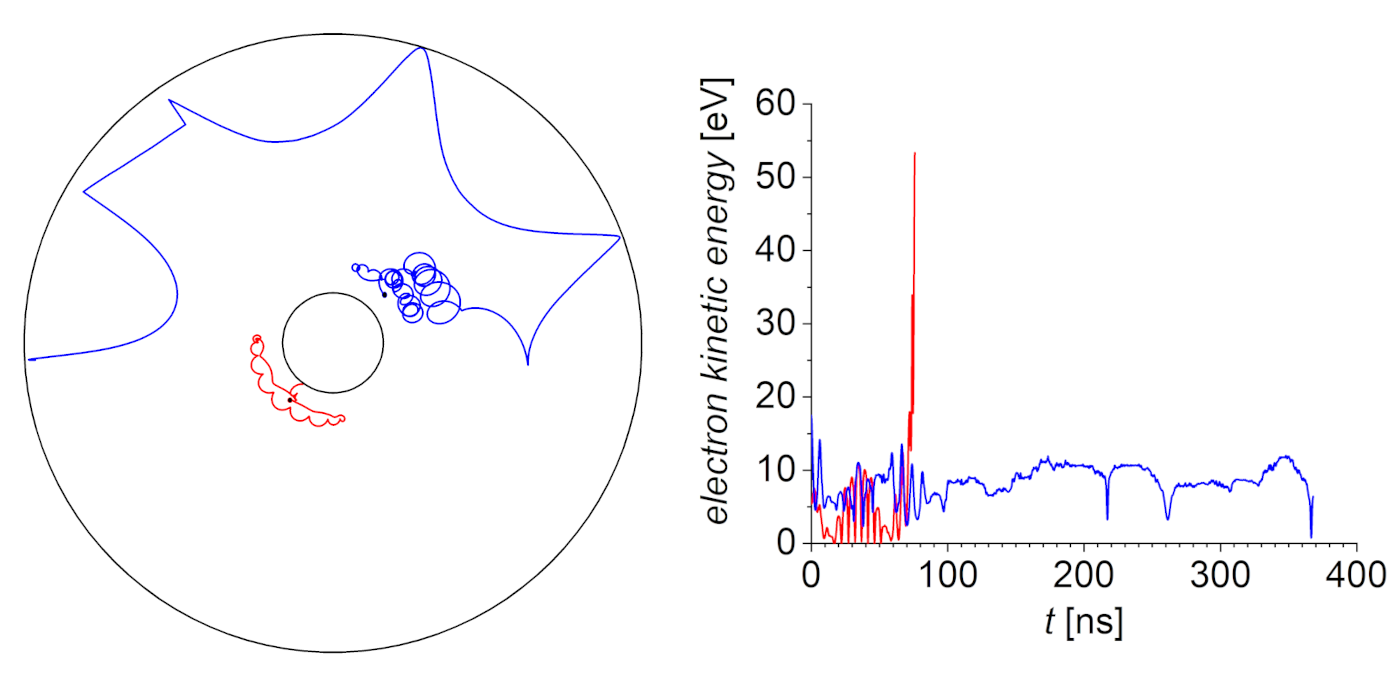}
\else
\fi
	\caption{Orbits (left) and kinetic energies (right) of two typical electrons with a similar initial radial location but different initial velocities (see description in the text). \FINAL}
	\label{Fig_el_orbits}
\end{figure}

The approximate solution gives only a qualitative picture. Due to the strong non-uniformity of particularly 
the 
electric field at the sheath edge, finite Larmor radius
effects play a role. 
\LB
In addition, the motion is strongly influenced by collisions.
Fig.~\ref{Fig_el_orbits} (left) shows the orbits of two electrons taken from the self-consistent simulation described in
section \ref{SectionWithMagneticField} below.\LB
Both electrons start near the electrode (marked by black circles) during sheath collapse.
The first electron (red) is initially trapped by the magnetic 
field.
It experiences a modulated electric field, which is positive
from $t= 0\,\mathrm{ns}$ to $t= 15\,\mathrm{ns}$, negative from $t=15\,\mathrm{ns}$ to $t=57\,\mathrm{ns}$,\LB and then positive again.
The corresponding azimuthal $\vec{E}\!\times\vec{B}$-drift results in a banana-like trajectory of the guiding center, the width of which is determined by the polarization drift.\LB
Fig.~\ref{Fig_el_orbits} (right) shows that
the kinetic energy is also modulated. 
At $t=73\,\mathrm{ns}$ and  $t=74\,\mathrm{ns}$,\LB the electron  undergoes elastic collisions which change its orbit and cause it to go to the electrode during the next phase. 
The other electron (blue) also remains initially trapped close to the electrode, despite collisions at 
$t=30\,\mathrm{ns}$ and $t=70\,\mathrm{ns}$. Between the elastic collisions,
\LB it experiences a strong electric field of changing polarity and changes its energy accordingly. 
A final collision at $t=100\,\mathrm{ns}$ scatters 
the particle near the grounded electrode where the magnetic field is weak. There it follows
an orbit with a very large gyroradius and is repeatedly reflected by the potential of the grounded sheath.


\pagebreak

\section{Kinetic description and velocity moments}

For a complete picture of the processes in the discharge, the dynamics of the particles 
(including collisions and wall interactions) must be coupled to the evolution of the fields.
In the low pressure regime of magnetrons, kinetic and non-local phenomena are expected,
and a kinetic approach is to be adopted. The particle sector is described by a collection of 
Boltzmann equations for the distribution functions $f_s = f_s(r,v_r,v_\theta,v_z,t)$, 
with $s= 1\dots N_\mathrm{s}$. 
(Our specific calculations use $s \in \{\mathrm{e}, \mathrm{i}\}$.)
For the streaming term on the left, the equations of motion (\ref{eq3_1}) are used, while the term on the
right describes the action of collisions:
\begin{equation}
\frac{\partial f_s}{\partial t} + v_r\frac{\partial f_s}{\partial r}
+\left(\frac{q_s}{m_s}\bigl(E_r + v_\theta B_z\bigr) +\frac{v_\theta^2}{r}\right)\frac{\partial f_s}{\partial v_r}
-\left(\frac{q_s}{m_s}v_r B_z +\frac{v_\theta v_r}{r}\right)\frac{\partial f_s}{\partial v_\theta} = 
C(f_s).
\label{BoltzmannEquation}
\end{equation}
Because of the small dimensions of the discharge and the relatively small electron density, 
the electrostatic approximation can be adopted. The radial electric field $E_r$ is the gradient of 
the electrostatic potential, $E_r = -\partial \Phi/\partial r$, and Poisson's equation can be written as 
\begin{equation}
-\varepsilon_0 \frac{1}{r}\frac{\partial}{\partial r} r \frac{\partial\Phi} {\partial r} =
   \sum_{s=1}^{N_\mathrm{s}} q_s \int f_s \, \mathrm{d}v_r\, \mathrm{d}v_\theta \, \mathrm{d}v_z. \label{eq3_1a}
\end{equation}

A direct solution of these equations with current computational resources is cumbersome. \LB
We thus employ ECCOPIC1S-M, 
a specialized member of our own ECCOPIC (Energy-\hspace{-0.05mm} and\hspace{-0.05mm} Charge-COnserving PIC)  suite of GPU-parallelized energy and charge conserving
particle-in-cell/ Monte Carlo collisions (PIC/MCC) codes \cite{Eremin2022}. 
ECCOPIC1S-M is an electro\-static 
1d3v code  for magnetized low pressure discharges in  Cartesian, cylindrical, or spherical geometry. 
\LB
It contains several innovations over standard PIC/MCC. 
The particle positions and potential\LB values are found self-consistently 
during a time step, 
using the Crank-Nicolson method inherent to  energy-conserving PIC/MCC schemes \cite{chen_2011}.
The algorithm relaxes the criterion for the onset of the finite grid instability which causes 
numerical heating in conventional PIC/MCC if the cell size is greater than the Debye length \cite{Barnes2021}. The accuracy of the orbit 
integration is controlled by an adaptive sub-stepping technique. 
An external network  with voltage source and blocking capacitor is fully integrated.
The collisions are 
evaluated with the null collision method \cite{vahedi_1995}
modified for GPUs \cite{mertmann_2011};  the cross sections are adopted 
from Phelps \cite{phelps_1999,phelps_1994}.\LB
As the plasma density is low, Coulomb collisions are  neglected.
The ion-induced secondary electron emission coefficient was assumed to be $\gamma= 0.1$ 
and the sticking coefficient $s=1$.  
\LB
For details on the implemented algorithms, see \cite{chen_2011,markidis_2011,Eremin2022}, 
for verification of the code family ECCOPIC and benchmarking, see \cite{turner_2013,charoy_2019,villafana_2021}.

 
The primary results of a PIC/MCC run are the
electric potential values on the grid and the particle 
positions as functions of time. Single particle trajectories can be recorded as well. \LB
More insight into the dynamics can be obtained by reconstructing 
the velocity moments of the  phase-space 
distribution function $f$. 
Defined abstractly as integrals over velocity space, \LB
they are realized in the simulation code as sums over all 
particles in the respective grid cells. \LB
To reduce noise,  an average is taken over 1000 RF periods after convergence has been reached.\LB
Focusing on the electrons 
and suppressing the index $\mathrm{e}$, we list the electron density
\begin{align}
	  n(r,t) = \int f(r,v_r,v_\theta,v_z,t)\, \mathrm{d}^3v,
\end{align}
the mean electron velocities in the directions $r$ and $\theta$,
\begin{align}
	  u_r(r,t) = \frac{1}{n}\int v_r f(r,v_r,v_\theta,v_z,t)\, \mathrm{d}^3v,\\[0.5ex]
		u_\theta(r,t) = \frac{1}{n}\int v_\theta f(r,v_r,v_\theta,v_z,t)\, \mathrm{d}^3v,
\end{align}
and the non-vanishing elements of the pressure tensor,
\begin{align}
&	  p_{rr}(r,t) = \int m(v_r-u_r)(v_r-u_r) f(r,v_r,v_\theta,v_z,t)\, \mathrm{d}^3v,\\[0.5ex]
&	  p_{r\theta}(r,t) = \int m(v_r-u_r)(v_\theta-u_\theta) f(r,v_r,v_\theta,v_z,t)\, \mathrm{d}^3v,\\[0.5ex]
&	  p_{\theta\theta}(r,t) = \int m(v_\theta-u_\theta)(v_\theta-u_\theta) f(r,v_r,v_\theta,v_z,t)\, \mathrm{d}^3v,\\[0.5ex]
&	  p_{zz}(r,t) = \int mv_z v_z f(r,v_r,v_\theta,v_z,t)\, \mathrm{d}^3v.
\end{align}
Furthermore, we define the following moments of the collision integral,
\begin{align}
	 &S = \int  C(f_s)\, \mathrm{d}^3v, \\[0.5ex]
	 &F_r = \int m v_r \, C(f_s)\, \mathrm{d}^3v, \\[0.5ex]
	 &F_\theta = \int mv_\theta \, C(f_s)\, \mathrm{d}^3v. 
\end{align}
Multiplying the kinetic equation with $1$, $m v_r$ and $m v_\theta$ and integrating 
over velocity space yields the particle balance and the momentum balances 
in the directions of $r$ and $\theta$:
\begin{align}
	&\frac{\partial n}{\partial t}
	+ \frac{1}{r}\frac{\partial}{\partial r}\left(r n u_{r}\right)
	 = 	S,\\[0.5ex] 
	&\frac{\partial}{\partial t}(m n u_r) 
	+ \frac{1}{r}\frac{\partial}{\partial r}\left(r\left(m n u_{r}^2
	            + p_{rr}\right)\right)
	 - \frac{1}{r}\left(p_{\theta\theta} +  m n u_{\theta}^2\right)  = - e n \left(E_r + u_\theta B_z \right) + 	F_r,\label{rMomentumBalance}\\[0.5ex] 
	&\frac{\partial}{\partial t}(m n u_{\theta}) %
+\frac{1}{r^2}\frac{\partial}{\partial r}\left(r^2 \left(m n u_{r}u_{\theta} +p_{r\theta}\right)\right) = e n u_{r} B_z  + F_\theta.\label{thetaMomentumBalance}
\end{align}

\pagebreak

\section{PIC/MCC results: Reference case without magnetic field \label{SectionWithoutMagneticField}}

\begin{figure}[h!]\centering
\iffigures	
   \includegraphics[width=\textwidth]{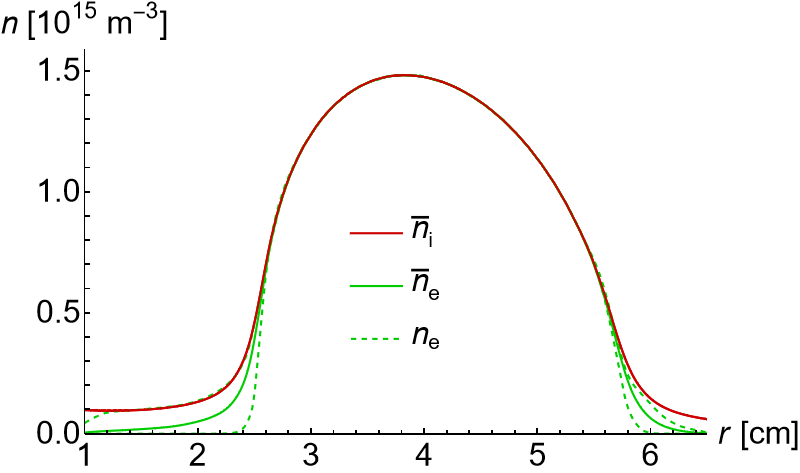}
\else
\fi
	\caption{RF period-averaged ion density  $\bar{n}_\mathrm{i}(r)$ (solid red) and electron density  $\bar{n}_\mathrm{e}(r)$ (solid green) of~the 
	unmagnetized reference discharge.
	The dashed green lines denote electron densities 
	$n_\mathrm{e}(r,t)$ at the sheath minimum (t = 0) and maximum (t = TRF/2), respectively.  \FINAL}
	\label{DensitiesUnmagnetizedCase}
\end{figure}

\begin{figure}[h!]\centering
\iffigures	
      \includegraphics[width=\textwidth]{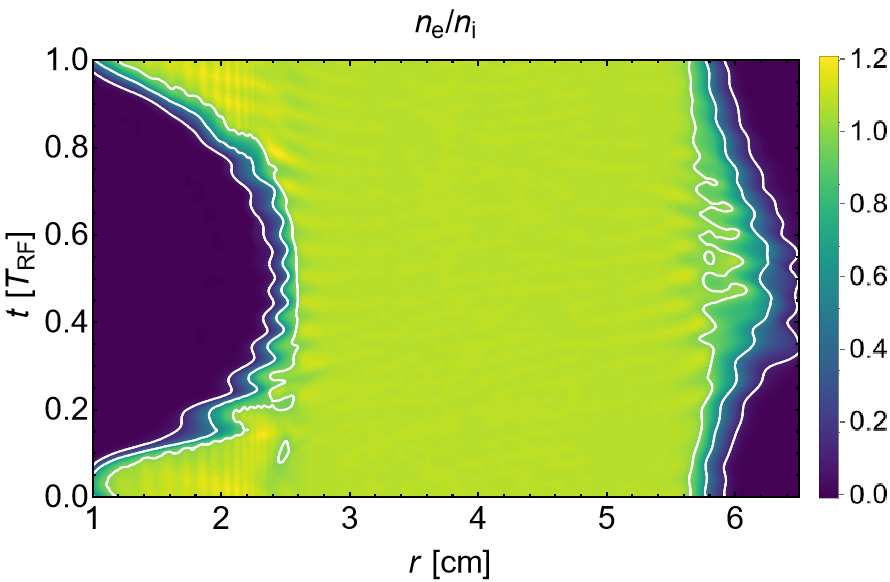}
\else
\fi
	\caption{Spatio-temporally resolved ratio of the electron density $n_\mathrm{e}(r,t)$ to the ion density $n_\mathrm{i}(r,t)$\LB
	for the unmagnetized reference discharge. The plasma-sheath 
	transition is indicated by the white  
	contours $n_\mathrm{e}/n_\mathrm{i}= 0.1,0.5,0.9$
	(from inside to outside) which will be used also in other figures. \FINAL}
	\label{RelativeElectronDensityUnmagnetizedCase}
\end{figure}

\begin{figure}[h!]\centering
\iffigures	
   \includegraphics[width=\textwidth]
              {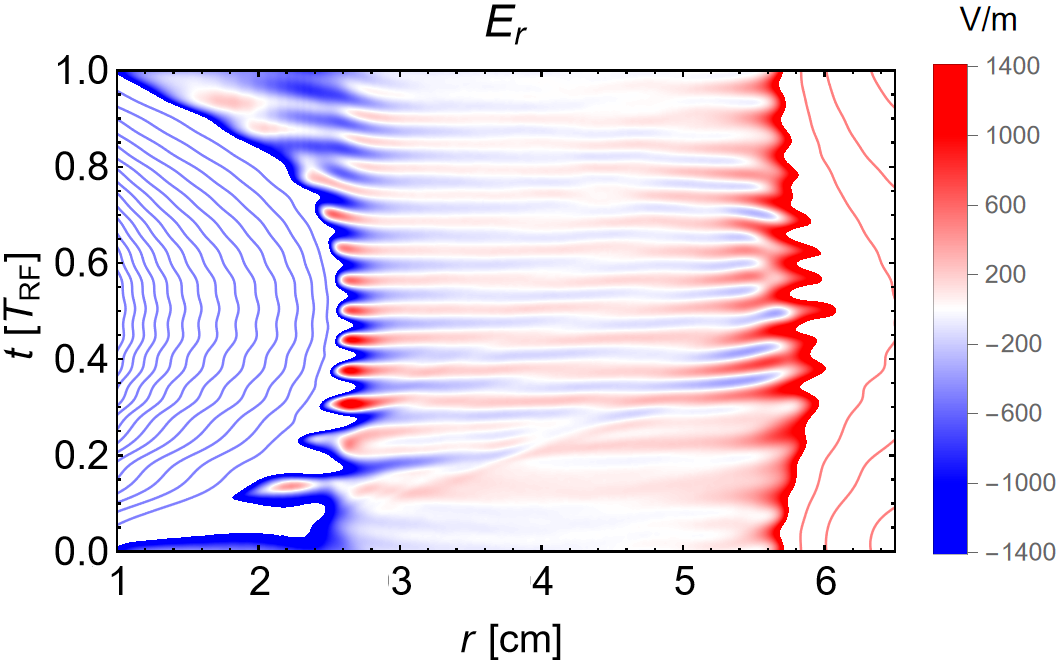}
\else
\fi
	\caption{Spatio-temporally resolved electric field $E_r(r,t)$ for the unmagnetized reference discharge.
	The color shading is scaled to resolve the weak fields in
	the plasma; the additional contour lines have a separation
	of $\Delta E = 5000\,\mathrm{V/m}$. \FINAL}
	\label{ElectricFieldUnmagnetizedCase}
\end{figure}

\begin{figure}[h!]\centering
\iffigures	
   \includegraphics[width=\textwidth]
   {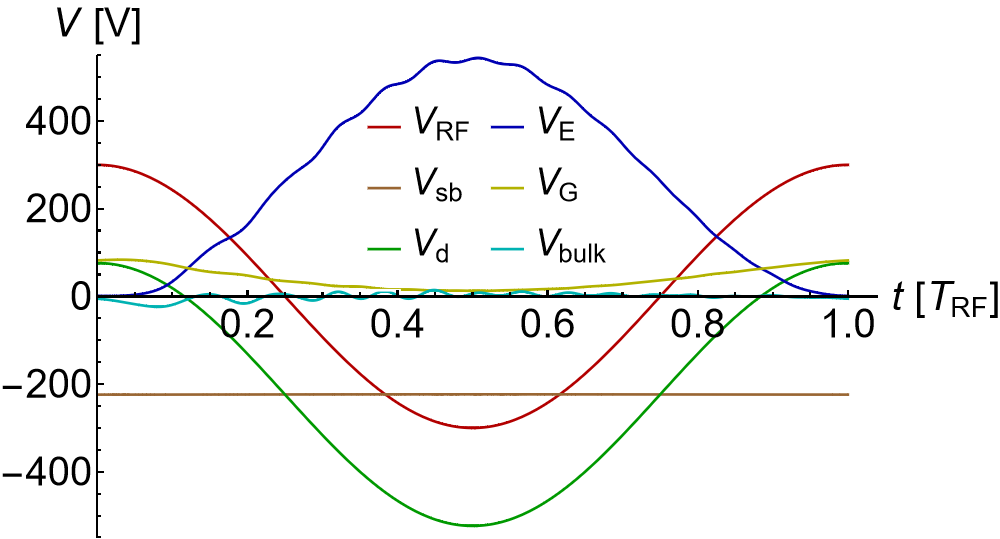}
\else
\fi
	\caption{Phase-resolved curves of the applied source voltage $V_\mathrm{RF}(t)$, the self bias voltage $V_\mathrm{sb}(t)$, the~total discharge voltage $V_\mathrm{d}(t)$,
	 the sheath voltages $V_\mathrm{E}(t)$ and $V_\mathrm{G}(t)$, and the voltage drop $V_\mathrm{bulk}(t)$
	 over the plasma bulk for the unmagnetized reference case.
	   \FINAL}
	\label{VoltagesUnmagnetizedCase}
\end{figure}

\begin{figure}[h!]\centering
\iffigures	
   \includegraphics[width=\textwidth]{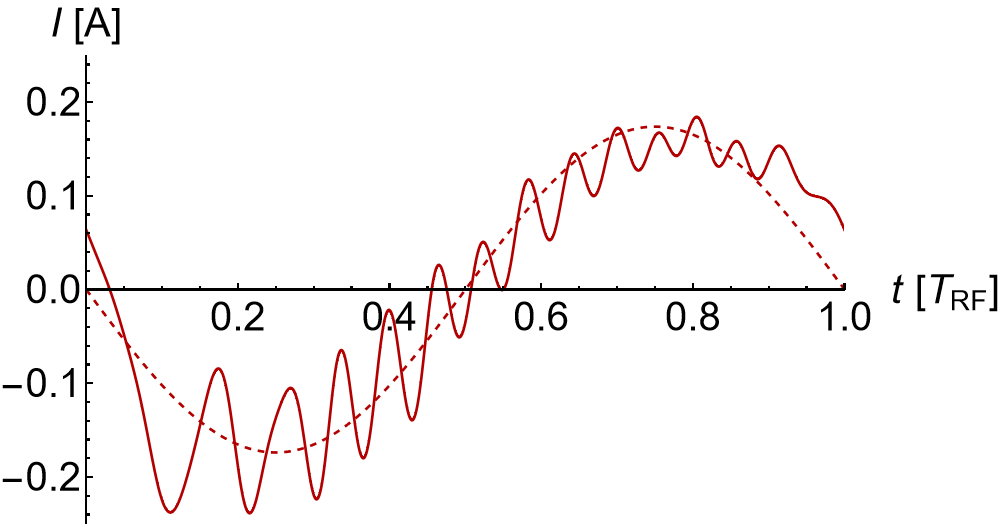}
\else
\fi
	\caption{
	Discharge current $I(t)$ of the unmagnetized case (solid) in comparison with the discharge current of the corresponding Drude model (dashed). \FINAL}
	\label{CurrentUnmagnetizedCase}
\end{figure}

\begin{figure}[h!]\centering
\iffigures	
   \includegraphics[width=\textwidth]{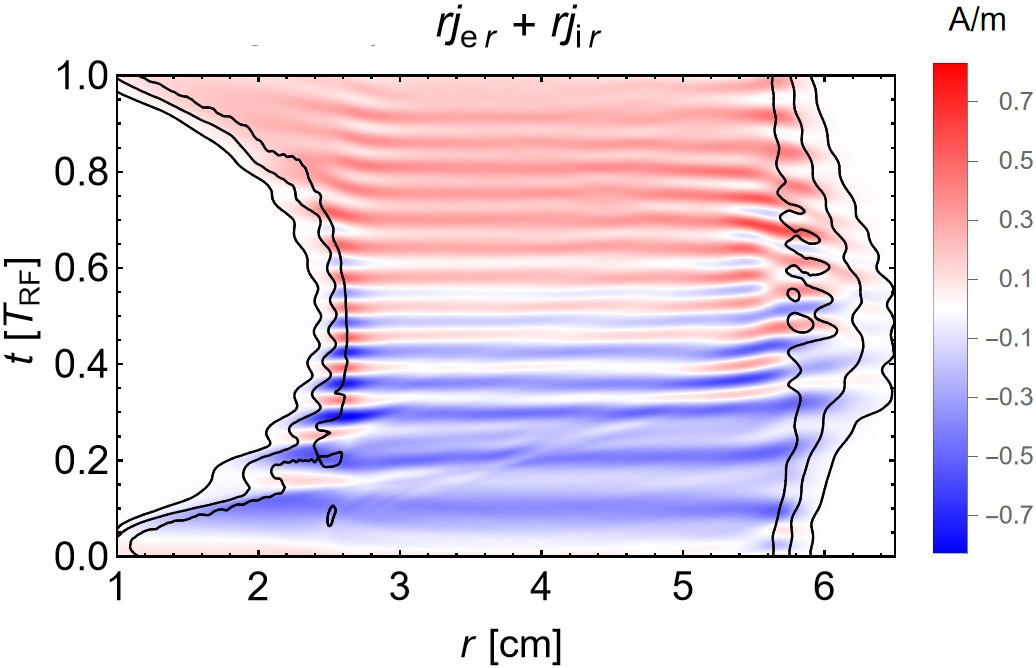}
\else
\fi
	\caption{
	Radially weighted particle current $rj_r(r,t)$ 	
	of the unmagnetized reference discharge. \FINAL}
	\label{WeightedParticleCurrentUnmagnetizedCase}
\end{figure}

\begin{figure}[h!]\centering
\iffigures	
   \includegraphics[width=\textwidth]{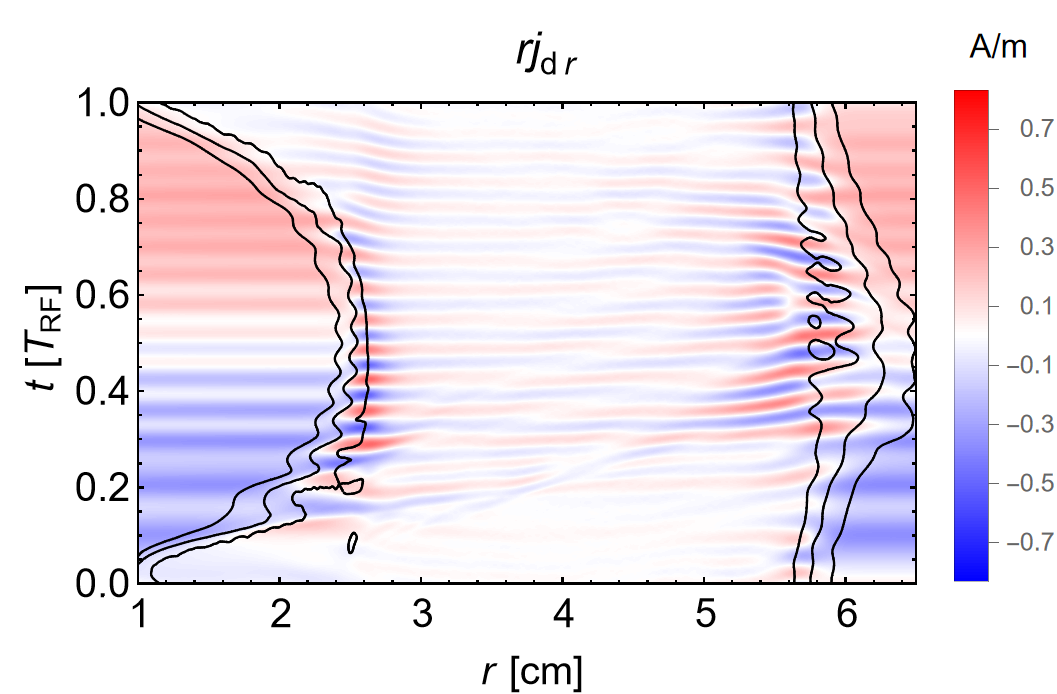}
\else
\fi
	\caption{
	Radially weighted displacement current $rj_{\mathrm{d}r}(r,t)$ of the 
	unmagnetized reference discharge.	\FINAL}
	\label{WeightedDisplacementCurrentUnmagnetizedCase}
\end{figure}

For reference, let us first consider the simpler unmagnetized case. 
It follows the logic of all RF-driven capacitive discharges:
The plasma self-organizes into the quasineutral 
bulk and two electron-depleted sheaths in front of the driven and grounded electrodes, respectively. \LB
 Fig.~\ref{DensitiesUnmagnetizedCase} shows the 
 phase-averaged densities of the electrons $\bar{n}_\mathrm{e}(r)$
 and the ions $\bar{n}_\mathrm{i}(r)$ and the\LB electron 
 densities $n_\mathrm{e}(r,t)$ at $t=0$ and $t=T_\mathrm{RF}/2$.
The peak density is $n = 1.5\!\times\! 10^{15}\,\mathrm{m}^{-3}$.
\LB
Fig.~\ref{RelativeElectronDensityUnmagnetizedCase} 
provides a map of the ratio $n_\mathrm{e}(r,t)/\bar{n}_\mathrm{i}(r)$,
Fig.~\ref{ElectricFieldUnmagnetizedCase} a map of the 
electric field $E_r(r,t)$. 
The electrons can follow the electric field, their plasma frequency $\omega_\mathrm{pe} = 2.2\!\times\! 10^9\,\mathrm{s}^{-1}$ (center)
or $\omega_\mathrm{pe} = 6.5\!\times\! 10^8\,\mathrm{s}^{-1}$ 
(electrode sheath) is larger than the radio frequency
$\omega_\mathrm{RF} = 8.5\!\times\! 10^7\,\mathrm{s}^{-1}$. 
\LB
The ions, in contrast,
  are almost unmodulated
and react only on the phase-averaged field;\LB
their plasma frequency is   
$\omega_\mathrm{pi} = 8.0\!\times\! 10^6\,\mathrm{s}^{-1}$ (center)
or $\omega_\mathrm{pi} = 2.4\!\times\! 10^6\,\mathrm{s}^{-1}$ 
(electrode sheath). 
\LB
Fig.~\ref{VoltagesUnmagnetizedCase} shows the applied voltage,
the discharge and sheaths voltages, and the bias voltage. 
The RF current $I(t)$ 
that flows through the discharge -- see Fig.~\ref{CurrentUnmagnetizedCase} -- 
is spatially constant, 
it is the  sum of the (small) ion current, the electron current, and  the displacement current. 
Figs.~\ref{WeightedParticleCurrentUnmagnetizedCase} and \ref{WeightedDisplacementCurrentUnmagnetizedCase} show spatio-temporally resolved maps of the 
components of this current. \LB
The factor $2\pi r$ reflects the cylindrical symmetry of the device, $H$ is its nominal height:
\begin{align}
 I(t) =  2\pi r H j(r,t)= 2\pi r H \left(e n_\mathrm{i} u_\mathrm{i} - e n_\mathrm{e} u_\mathrm{e} + 
\varepsilon_0 \frac{\partial E_r}{\partial t}\right).   
\end{align}

\begin{figure}[h!]\centering
\iffigures	
   \includegraphics[width=\textwidth]{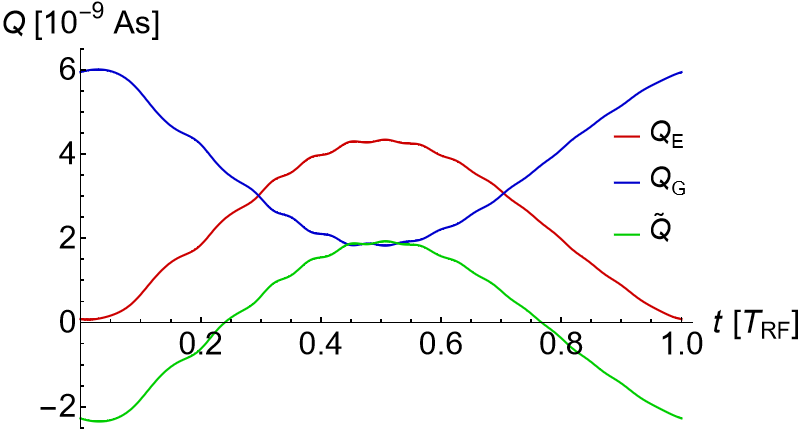} 
\else
\fi
	\caption{Charges of the electrode sheath $Q_\mathrm{E}(t)$ and the ground 
	sheath $Q_\mathrm{G}(t)$ together with the average-free integral $\tilde{Q}(t)$
	of the current $I(t)$ for the unmagnetized reference case. \FINAL}
	\label{SheathChargesUnmagnetizedCase}
\end{figure}

Due to the cylindrical device geometry, the current densities at 
the driven and the grounded electrode  
scale like $j_\mathrm{E}\!:\!j_\mathrm{G} = R_\mathrm{G}\!:\!R_\mathrm{E} =  6.5\!:\!1$.
This makes the discharge strongly asymmetric; 
the inner sheath is much wider 
($\bar{s}_\mathrm{E}= 1.1\,\mathrm{cm}$) than the grounded outer sheath ($\bar{s}_\mathrm{G}= 0.56\,\mathrm{cm}$). \LB
To understand the electric behavior of the discharge, it is useful to 
consult a simplified model which analyzes
the evolution of the charges in the driven and grounded sheath,
\begin{align}
    Q_{E} = H \int_{R_\mathrm{E}}^{R_\mathrm{C}} e(n_\mathrm{i}-n_\mathrm{e})\, 2\pi r \, \mathrm{d}r, \\[0.5ex]
     Q_\mathrm{G} = H \int_{R_\mathrm{C}}^{R_\mathrm{G}} e(n_\mathrm{i}-n_\mathrm{e}) \, 2\pi r \, \mathrm{d}r.
\end{align}
The RF current modulates the charges of the two sheaths. Neglecting small 
terms related to charge carrier losses to the electrodes, we write as follows,
where the charge quantity $\tilde{Q}(t)$ \LB is the phase-average-free 
integral of $-I(t)$, see Fig.~\ref{SheathChargesUnmagnetizedCase}: 
\begin{align}
     \frac{\mathrm{d}Q_\mathrm{E}}{\mathrm{d}t} = 
     - \frac{\mathrm{d}Q_\mathrm{G}}{\mathrm{d}t}= -I(t)=
      \frac{\mathrm{d}\tilde{Q}}{\mathrm{d}t}.
\end{align}

\begin{figure}[h!]\centering
\iffigures	
   \includegraphics[width=\textwidth]{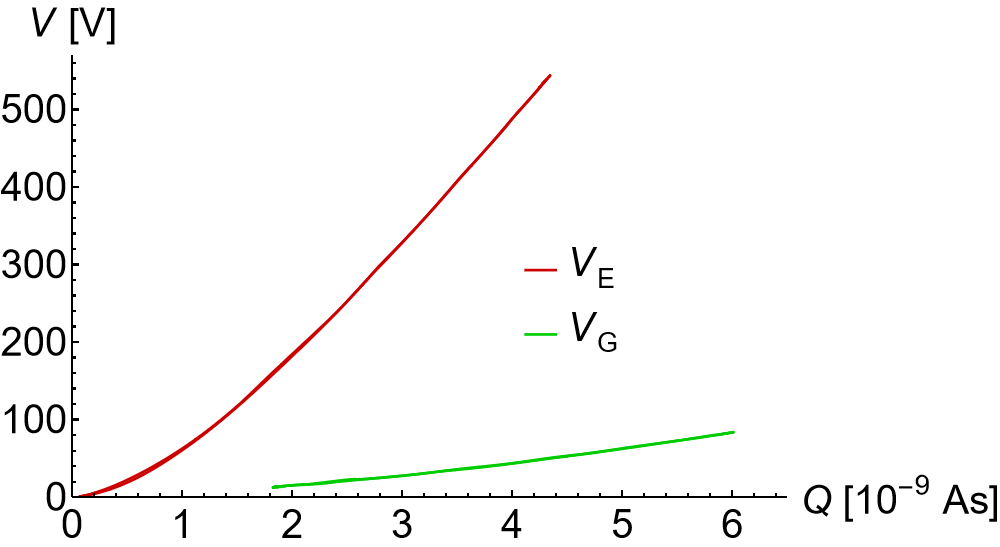}
\else
\fi
	\caption{Charge-voltage relation $V_\mathrm{E}(Q)$ of the 
	electrode sheath and
	$V_\mathrm{G}(Q)$ the ground sheath for the unmagnetized reference case.
	\FINAL}
	\label{VQUnmagnetized}
\end{figure}

The voltage drop over the discharge $V(t)$ is equal to the sum of the applied RF voltage
$V_\mathrm{RF}(t)$ \LB 
and the nearly constant self-bias $V_\mathrm{sb} = -224\,\mathrm{V}$. 
Assuming that the sheath voltages $V_\mathrm{E}$ and  $V_\mathrm{G}$ 
are functions of $Q_\mathrm{E}$ and  $Q_\mathrm{G}$,
respectively (Fig.~\ref{VQUnmagnetized}), and neglecting 
all other electric fields,
\LB
a voltage balance can be formulated.
Here, the offset charges $\bar{Q}_\mathrm{E}$, $\bar{Q}_\mathrm{G}$ and 
the self bias $V_\mathrm{sb}$\LB are constants which self-adjust so that 
the floating condition holds:
\begin{align}
    - V_\mathrm{E}\left(\bar{Q}_\mathrm{E}+\tilde{Q}(t)\right) + V_\mathrm{G}\left(\bar{Q}_\mathrm{G}-\tilde{Q}(t)\right) 
    -  V_\mathrm{sb}= V_\mathrm{RF}\cos(\omega t). 
    \label{VoltageBalance}
\end{align}
This quasistatic model explains much of the electrical behavior: 
The sheath voltages 
and the bias voltage 
are determined by the voltage $V_\mathrm{RF}(t)$.
The periodic charging and discharging of the sheaths gives rise to the discharge current $I(t)$. 
However, the model over-idealizes:
\LB
Its predicts that the charges and voltages are \textit{exact} functions of
the RF voltage $V_\mathrm{RF}(t)$,\LB and that the discharge current $I(t)$, 
being a derivative of $\tilde{Q}(t)$, has an \textit{exact} 90° phase shift. \LB
This would imply zero net heating. 
(See the appendix of \cite{SchulzeDonkoLafleurWilczekBrinkmann2018}
for details on this argument.)

In reality, and in the results of the PIC/MCC simulation, 
the current-voltage phase angle is not exactly 90° but very close to it,
\begin{align}
    \cos \varphi = \frac{1}{T} \int_0^T I V \,\mathrm{d}t\Bigg/
               \sqrt{\frac{1}{T} \int_0^T I^2 \,\mathrm{d}t\,\frac{1}{T} \int_0^T V^2 \,\mathrm{d}t}
       = 0.0439 = \cos 87.5\text{°},
\end{align}
and the phase-averaged dissipation is not zero but 
\begin{align}
    \bar{P} = \ \frac{1}{T} \int_0^T I V \,\mathrm{d}t
    = 1.22\,\mathrm{W}.
\end{align}
Responsible for the non-vanishing heating are, in fact, the small 
``other electric fields'' that were neglected in the voltage balance \eqref{VoltageBalance}. 
These small fields are needed to locally drive the current density $j(r,t)$
which corresponds to the globally determined discharge current $I(t)$. \LB
(They are also not well represented in the cold plasma model whose predictions of $\phi=89.96$° for the phase angle and $\bar{P}=0.018\,\mathrm{W}$ 
for the Ohmically dissipated power are unrealistic. \LB
However, the cold plasma model explains well the current oscillations 
related to the PSR;\LB the predicted value of $\omega_\mathrm{PSR} = 1.29\,\times\,10^9\,\mathrm{s}^{- 1}$ agrees with the outcome of the
simulations.)   \LB
%
In the vicinity of the sheaths, where the local plasma frequency
$\omega_\mathrm{pe}(r)$ is close to $\omega_\mathrm{PSR}$,\LB
also Langmuir oscillations occur \cite{Wilczek2020}; see the phase-resolved maps of the electric field 
$E_r(r,t)$,\LB the particle current $j_{r}(r,t)$, and the 
displacement current $j_\mathrm{d}(r,t)$. The hallmark of Langmuir 
oscillations is a 180° degree phase shift between $j_{r}$
and $j_\mathrm{d}$.

\begin{figure}[h!]\centering
\iffigures	
   \includegraphics[width=17cm]{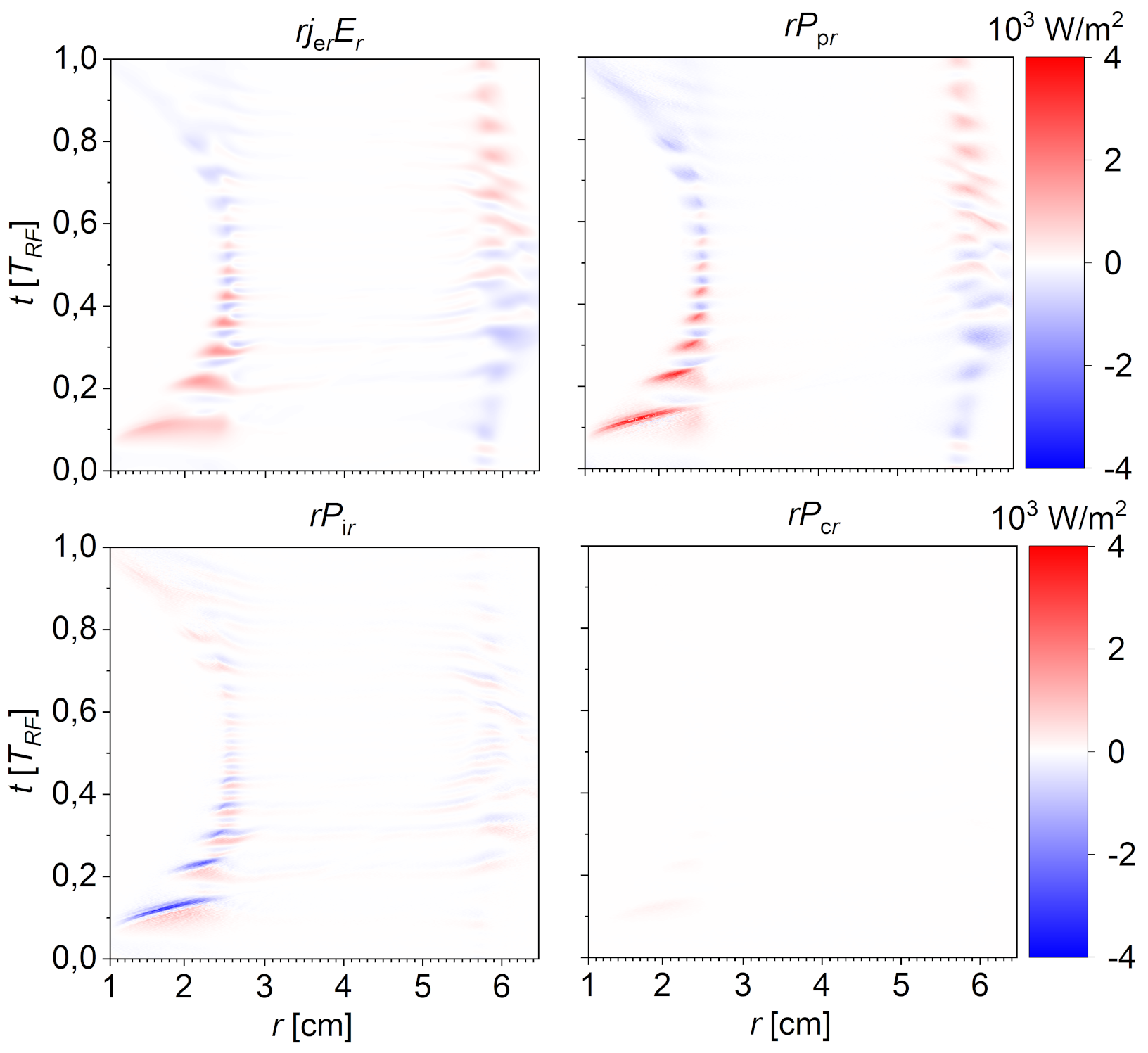} 
\else
\fi
	\caption{The radially weighted spatio-temporally resolved total power transfer $r P_{\mathrm{e}r}$ from the electric field to electrons in the radial direction and its 
constituents ($P_{\mathrm{p}r}$, $P_{\mathrm{c}r}$, $P_{\mathrm{i}r}$) from Eq.(\ref{PowerbalanceUnmagnetizedCase}) for the unmagnetized reference discharge. 
}
	\label{Fig_pow_abs_unmagn}
\end{figure}

Of the power $\bar{P}=1.22\,\mathrm{W}$ provided by the RF source,
$\bar{P}_\mathrm{i}=0.91\,\mathrm{W}$ are absorbed by the ions
\LB
and $\bar{P}_\mathrm{e}=0.31\,\mathrm{W}$ by the electrons.
The energy transfer to the ions takes place in the sheaths and can be described
in terms of the period-averaged field $\bar{E}_r$.
The energy exchange between the electric field and  the electrons, however, is a complex spatio-temporal phenomenon, \LB involving both the RF modulation 
and the self-excited PSR. Insight is gained from a velocity moment analysis of the kinetic 
equation \cite{surendra_1993,SchulzeDonkoLafleurWilczekBrinkmann2018}.
We solve the momentum balance \eqref{rMomentumBalance}
for the electric field which then can be separated into the inertia field, 
the pressure field,\LB and the collisional field:
\begin{align}
\label{EanalysisUnmagnetized}
 E_r &=-\frac{1}{e n}\left( \frac{\partial}{\partial t}(m n u_r)
            + \frac{1}{r}
\frac{\partial}{\partial r}\left(r m n u_r^2 \right)-
\frac{m n u_\theta^2}{r}\right)- \frac{1}{r e n}
\left(\frac{\partial}{\partial r}\left(r p_{rr} \right)- p_{\theta\theta}\right)
    + \frac{\dot{\pi}_r}{e n} \\[0.5ex] \nonumber
    & \equiv E_{\mathrm{i}r}+E_{\mathrm{p}r} + E_{\mathrm{c}r}.
\end{align}
%
%
Multiplying by $r j_r = - r e n u_r $, we establish the mechanical energy balance
and obtain a split of the radially weighted electron power dissipation $r P_{\mathrm{e}r}$, see Fig.~\ref{Fig_pow_abs_unmagn},
%
\begin{align}  
\label{PowerbalanceUnmagnetizedCase}
 r P_{\mathrm{e}r}= r j_e E_r  = & \;
	r m u_r \left(\frac{\partial}{\partial t}(n u_r)
	+  \frac{1}{r}\frac{\partial}{\partial r}\left(r n u_{r}^2
	            \right) 
	            -  \frac{n u_\theta u_\theta}{r} \right)
	+ u_r \left(\frac{\partial}{\partial r}\left(r p_{rr}\right)
	 - r  \frac{p_{\theta\theta}}{r} \right) - r u_r	\dot{\pi}_r  \nonumber\\[0.5ex]  
	 \equiv &  \; r P_{\mathrm{i}r}  + r P_{\mathrm{p}r} + r P_{\mathrm{c}r}. 
\end{align}
There is almost no electron heating in the plasma bulk; virtually all RF power is absorbed in the 
temporarily quasi\-neutral sections of the sheaths.
All the heating processes discussed in section \ref{Introduction} are present, 
but are of different importance. ``Classical'' Ohmic heating is weak: 
\LB
The collisional term $r P_\mathrm{c}$ contributes only 25\,\%
to  $r P_\mathrm{e}$, not surprising at the small 
gas pressure of only $p = 0.5\,\mathrm{Pa}$. 
The inertia term
$r P_\mathrm{i}$ can be momentarily large but is small on average.\LB
(This reflects that inertia forces are often non-dissipative: 
Electrons accelerated during one half phase of PSR osillations can be decelerated in the other and feed their energy back.)
The main contribution is thus embodied in the pressure term $P_{\mathrm{p}r}$. 

\begin{figure}[h!]\centering
\iffigures	
   \includegraphics[width=\textwidth]{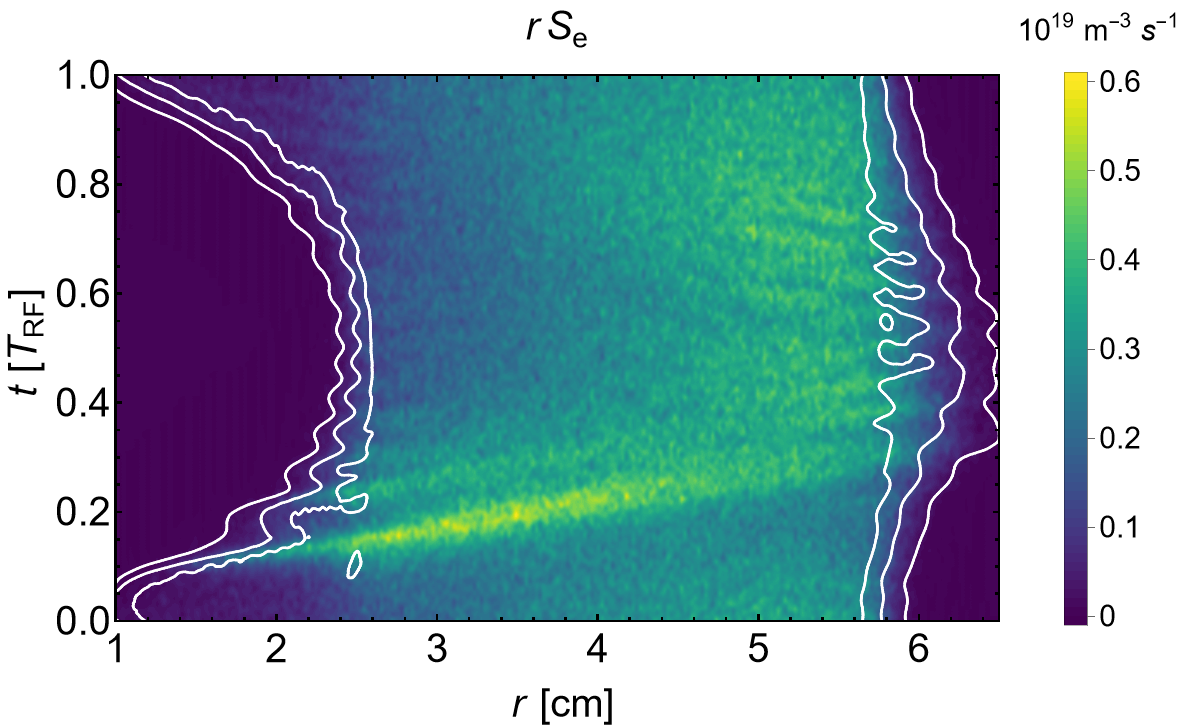}
\else
\fi
	\caption{The radially weighted, spatio-temporally resolved            profile of the ionization term for the 
	         unmagnetized reference discharge.
\FINAL
}
	\label{Fig_en_el_unmagn}
\end{figure}

A kinetic analysis can refine this picture. Immediately after sheath collapse, 
in the interval $t=0.02\,T_\mathrm{RF}$ to  $t=0.14\,T_\mathrm{RF}$, 
the electrode sheath expands with a speed of 
$u_\mathrm{s} \approx 2\times 10^6\,\mathrm{m/s}$, \LB
This high speed arises from a synergy 
 of the RF modulation and the PSR oscillation excited 
 at the moment of sheath collapse \cite{czarnetzki_2006}.
The fast moving sheath reflects the incoming electrons according to 
$v \to 2 u_\mathrm{s}-v$.
A  group of energetic electrons forms, which, further accelerated by the ambipolar field of the presheath, propagate beam-like towards the grounded electrode. \LB
This can be seen in  Fig.~\ref{Fig_en_el_unmagn}
which displays the radially weighted, spatio-temporally resolved profile of the ionization term.
Before the grounded electrode, 
the energetic beam is stopped by the ambipolar field and by the retreating sheath,
giving rise to a negative power density.
\LB 
However, before reaching that region, some of them 
have lost their directed energy already by collisions in the bulk, 
so that even if the rest is fully decelerated by the opposite sheath, 
the period-averaged 
energy transfer rate is positive. 
In the second half of the RF period,  the sheaths switch roles,
but with less effect on the heating process, since the sheath before the grounded electrode is less strongly modulated.
Another source of temporal asymmetry can be attributed to different electron temperatures during sheath expansion and collapse,  also resulting in a positive period-averaged power absorption 
\cite{SchulzeDonkoDerzsiKorolovSchuengel2015,SchulzeDonkoLafleurWilczekBrinkmann2018}.   
Heating is therefore substantial in two situations: When there is a violent acceleration of a significant fraction of particles at the sheath edge,
or at the sheath-presheath boundary where a significant 
temporally asymmetric ambipolar field is present
\cite{SchulzeDonkoDerzsiKorolovSchuengel2015}.

%
%

\pagebreak

\section{Case with magnetic field \label{SectionWithMagneticField}}

\begin{figure}[h!]\centering
\iffigures	
   \includegraphics[width=15cm]{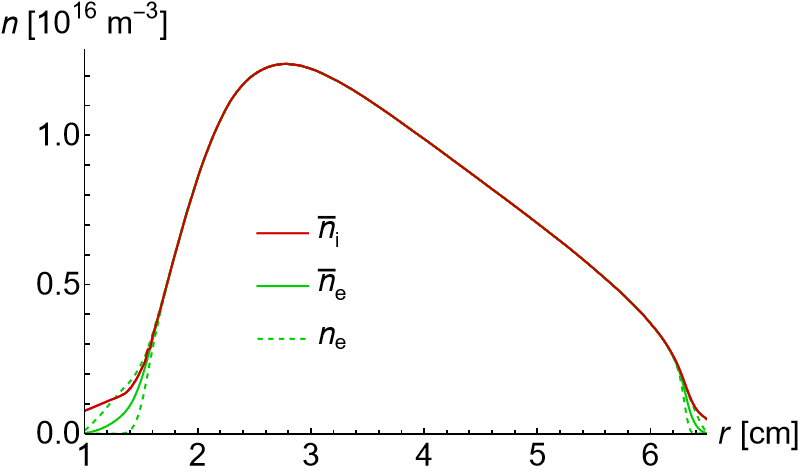} 
\else
\fi
	\caption{RF Period-averaged ion density $\bar{n}_\mathrm{i}(r)$ (solid red) 
	         and electron density $\bar{n}_\mathrm{e}(r)$ (solid green) \LB
	         of the magnetized discharge. The dashed green lines represent the electron
	         densities $n_\mathrm{e}(r,t)$  at the moments of electrode sheath minimum ($t=0$) 
	         and maximum ($t=T_\mathrm{RF}/2$), respectively.
	         \FINAL}
	\label{ParticleDensitiesMagnetized}
\end{figure}

We now turn to the more complex magnetized case. (Fig.~\ref{DischargeSchemeDrawing} gives
a true-to-scale schematic.)\LB
Also this discharge self-organizes into a 
quasineutral bulk and two electron-depleted sheaths in front of the driven and grounded 
electrodes, respectively.
Its plasma density, however,\LB is $7.5$ times higher, the peak density is 
$n_\mathrm{max} = 1.24\!\times\! 10^{16}\,\mathrm{m}^{-1}$.
The sheaths are thinner by nearly a factor of four, 
$\bar{s}_{\mathrm{E}}= 0.32\,\mathrm{cm}$ and $\bar{s}_{\mathrm{G}}= 0.14\,\mathrm{cm}$. 
Fig.~\ref{ParticleDensitiesMagnetized} shows the phase-averaged densities 
$\bar{n}_\mathrm{i}(r)$ and $\bar{n}_\mathrm{e}(r)$ 
along with snapshots of 
$n_\mathrm{e}(r,t)$ at the phases $t=0$ and $t=T_\mathrm{RF}/2$.

\begin{figure}[h!]\centering
\iffigures	
      \includegraphics[width=\textwidth]
      {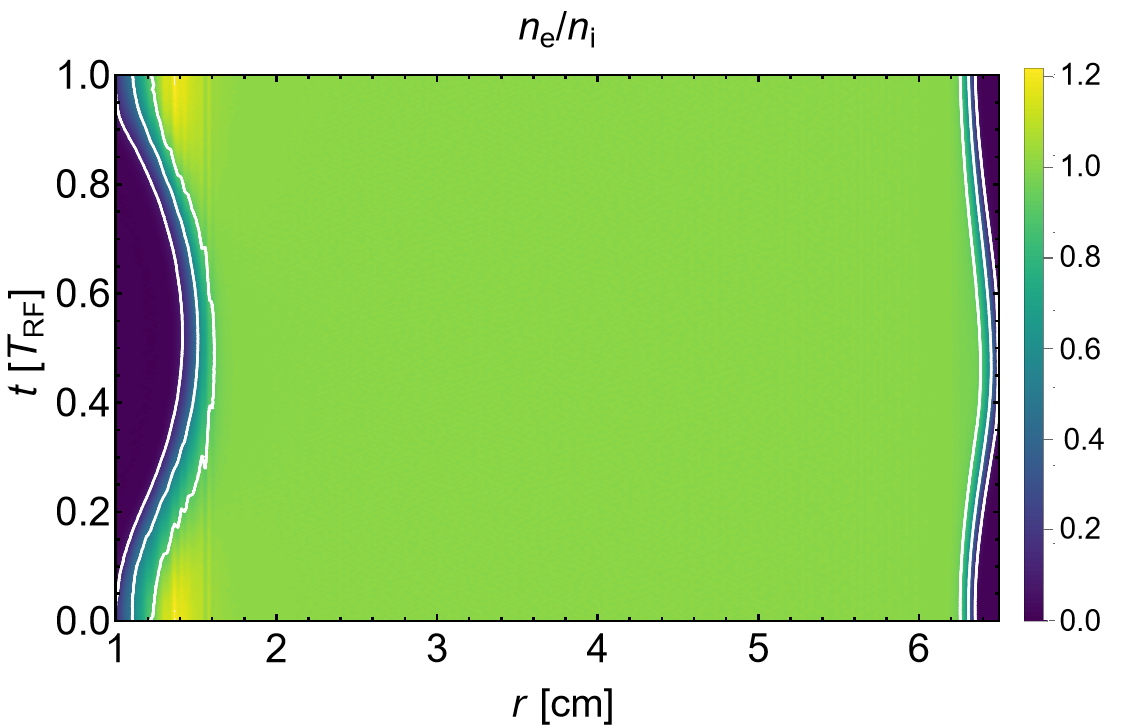}
\else
\fi
	\caption{Spatio-temporally resolved ratio of the electron density $n_\mathrm{e}(r,t)$ to the ion density $n_\mathrm{i}(r,t)$\LB
	for the magnetized discharge. The plasma-sheath transition is indicated by the white   
	contours $n_\mathrm{e}/n_\mathrm{i}= 0.9,0.5,0.1$ (from inside to outside)
	which will be used also in other figures. 
	\FINAL}
	\label{RelativeElectronDensityMagnetizedCase}
\end{figure}

\begin{figure}[h!]\centering
\iffigures	
   \includegraphics[width=\textwidth]
              {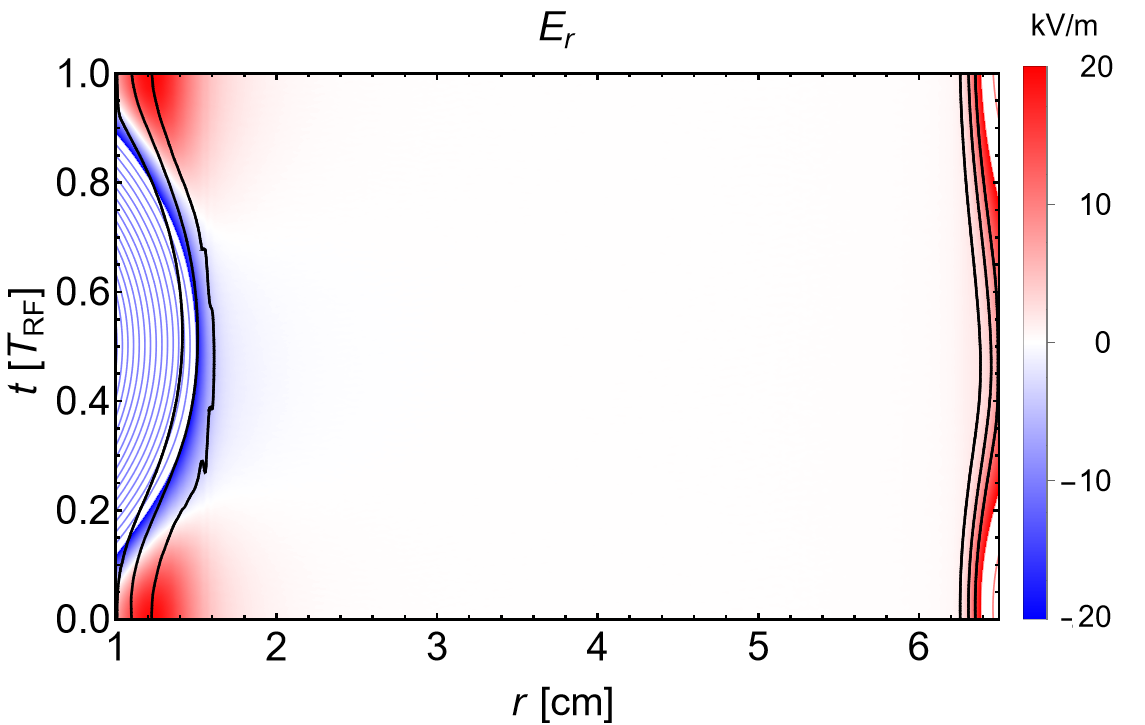}
\else
\fi
	\caption{Spatio-temporally resolved electric field $E_r(r,t)$ for the magnetized magnetron discharge.
	The color shading is scaled to resolve the weak fields in
	the temporarily quasi-neutral regions; the~additional contour lines have a separation
	of $\Delta E = 20\,\mathrm{kV/m}$. \FINAL}
	\label{ElectricFieldMagneticCase}
\end{figure}

Fig.~\ref{RelativeElectronDensityMagnetizedCase} 
provides a map of the ratio $n_\mathrm{e}(r,t)/\bar{n}_\mathrm{i}(r)$,
Fig.~\ref{ElectricFieldMagneticCase} a map of the 
electric field $E_r(r,t)$. \LB
Again, the electrons can follow the field,
their plasma frequency is $\omega_\mathrm{pe} = 5.7\!\times\! 10^9\,\mathrm{s}^{-1}$ 
(center) \LB or $\omega_\mathrm{pe} = 2.2\!\times\! 10^9\,\mathrm{s}^{-1}$ 
(electrode sheath), while the ion component is essentially unmodulated,  
$\omega_\mathrm{pi} = 2.1\!\times\! 10^7\,\mathrm{s}^{-1}$ (center) 
or $\omega_\mathrm{pi} = 8.1\!\times\! 10^6\,\mathrm{s}^{-1}$ 
(electrode sheath). 

\begin{figure}[h!]\centering
\iffigures	
   \includegraphics[width=\textwidth]
   {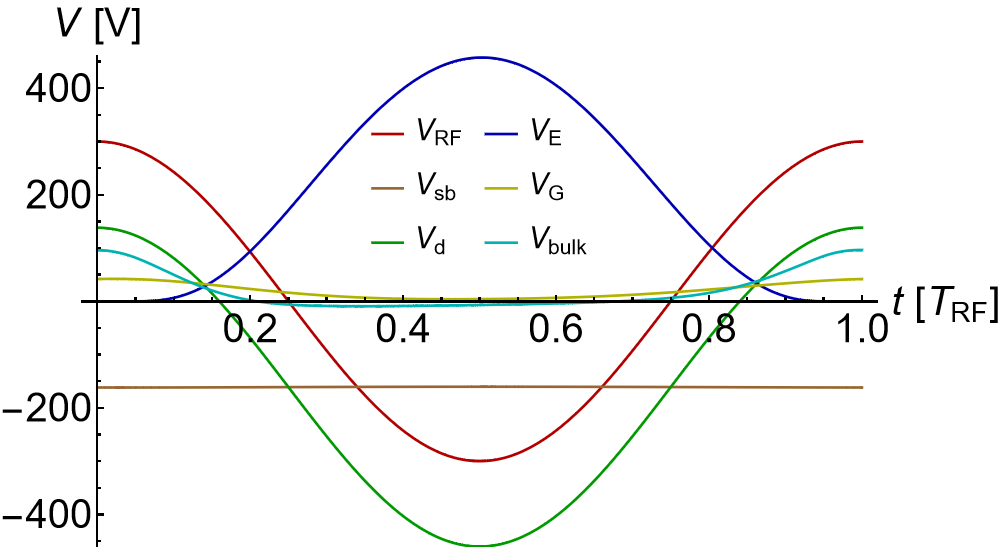}
\else
\fi
	\caption{Phase-resolved curves of the applied source voltage $V_\mathrm{RF}(t)$, the self bias voltage $V_\mathrm{sb}(t)$, the~total discharge voltage $V_\mathrm{d}(t)$,
	 the sheath voltages $V_\mathrm{E}(t)$ and $V_\mathrm{G}(t)$, and the voltage drop $V_\mathrm{bulk}(t)$
	 over the plasma bulk for the magnetized case. \FINAL}
	\label{VoltagesMagnetizedCaseNEW}
\end{figure}
\begin{figure}[h!]\centering
\iffigures	
   \includegraphics[width=\textwidth]{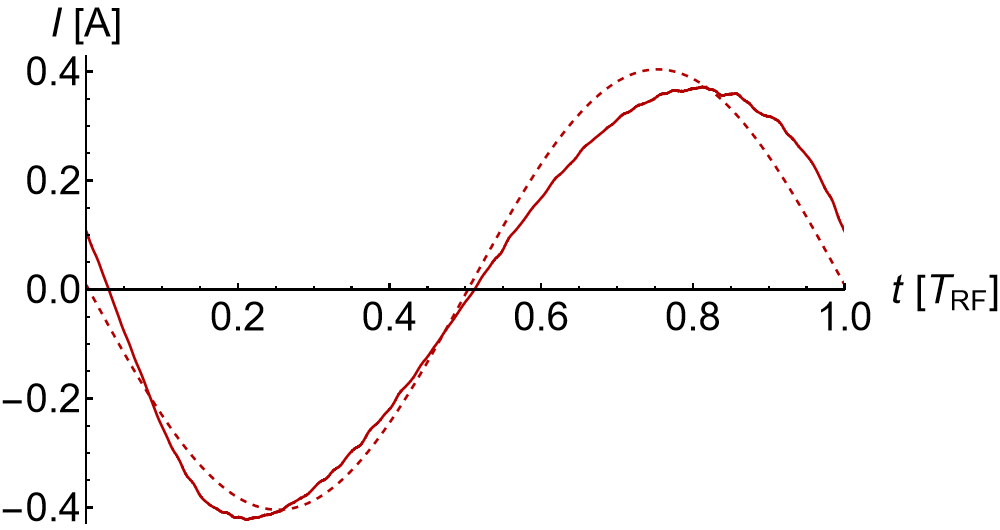}
\else
\fi
	\caption{Discharge current $I(t)$ of the magnetized case (solid) in comparison with the discharge current of the corresponding Drude model (dashed). \FINAL}
	\label{CurrentMagnetizedCase}
\end{figure}
\begin{figure}[h!]\centering
\iffigures	
   \includegraphics[width=\textwidth]{
   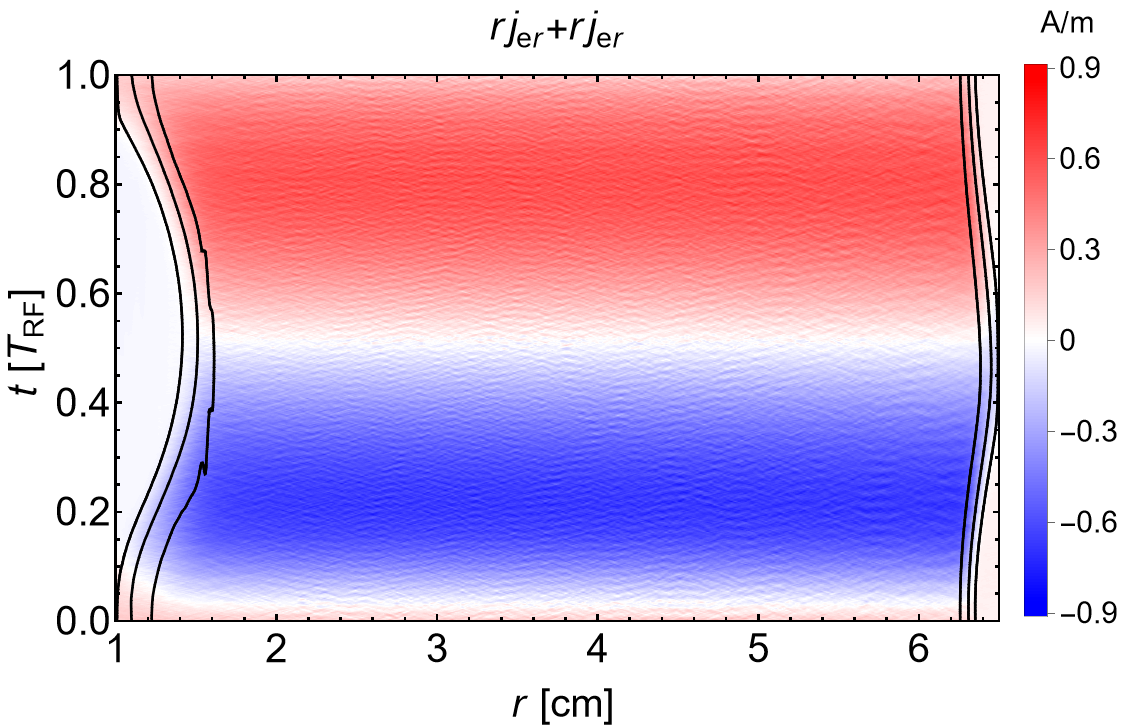}
\else
\fi
	\caption{
	Radially weighted particle current $rj_r(r,t)$ in the 
	magnetized case.
	\FINAL}
	\label{WeightedParticleCurrentMagnetizedCase}
\end{figure}
\begin{figure}[h!]\centering
\iffigures	
   \includegraphics[width=\textwidth]{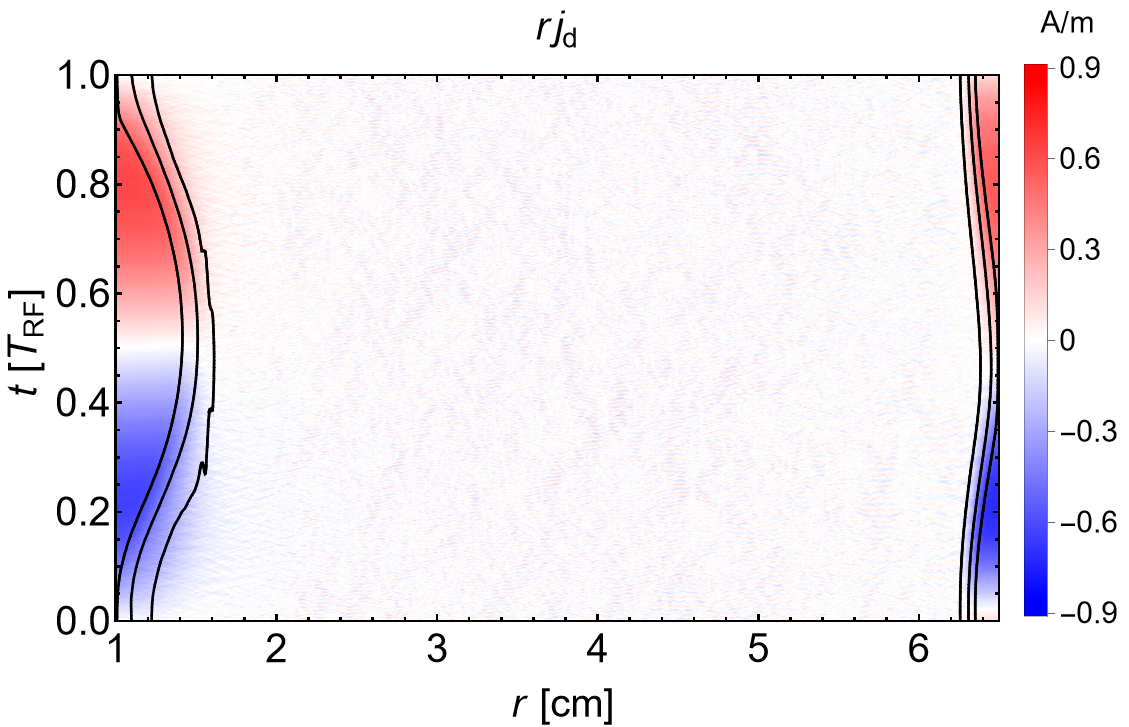}
\else
\fi
	\caption{
	Radially weighted displacement current for the magnetized case.
	\FINAL}
	\label{WeightedDisplacementCurrentMagnetizedCase}
\end{figure}
\begin{figure}[h!]\centering
\iffigures	
   \includegraphics[width=\textwidth]
                 {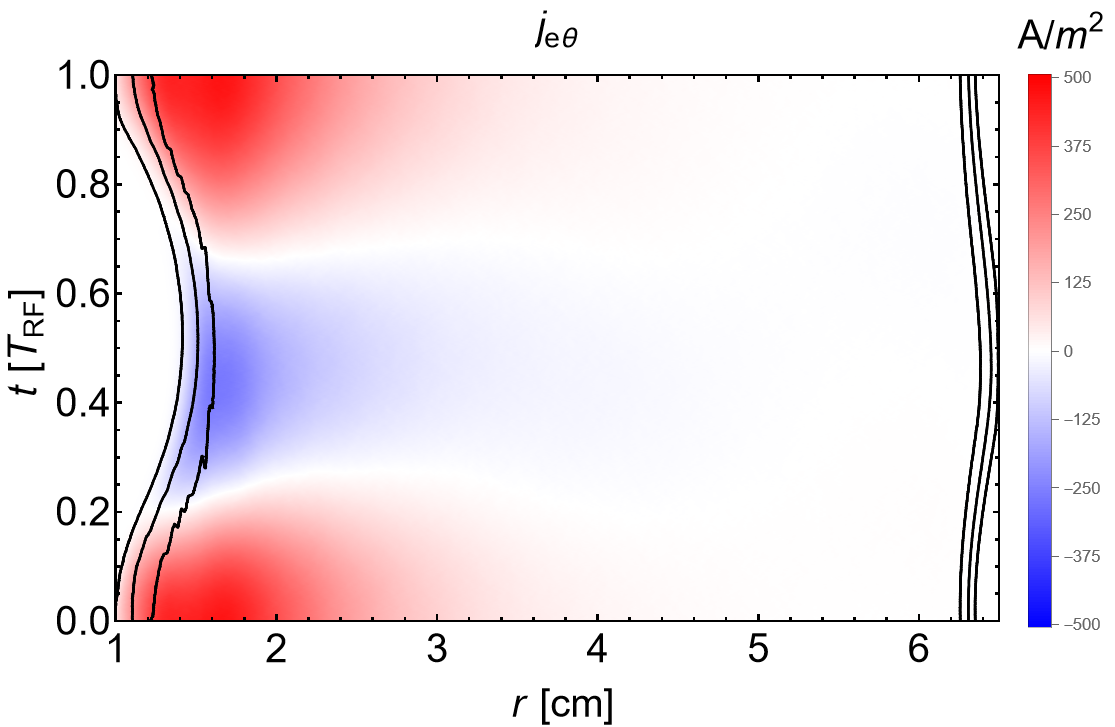}
\else
\fi
	\caption{
	Azimuthal electron current density $j_{\mathrm{e}\theta}(r,t)$ for the magnetized case. 
	\FINAL}
	\label{jthetaCurrent}
\end{figure}
\begin{figure}[h!]\centering
\iffigures	
   \includegraphics[width=\textwidth]{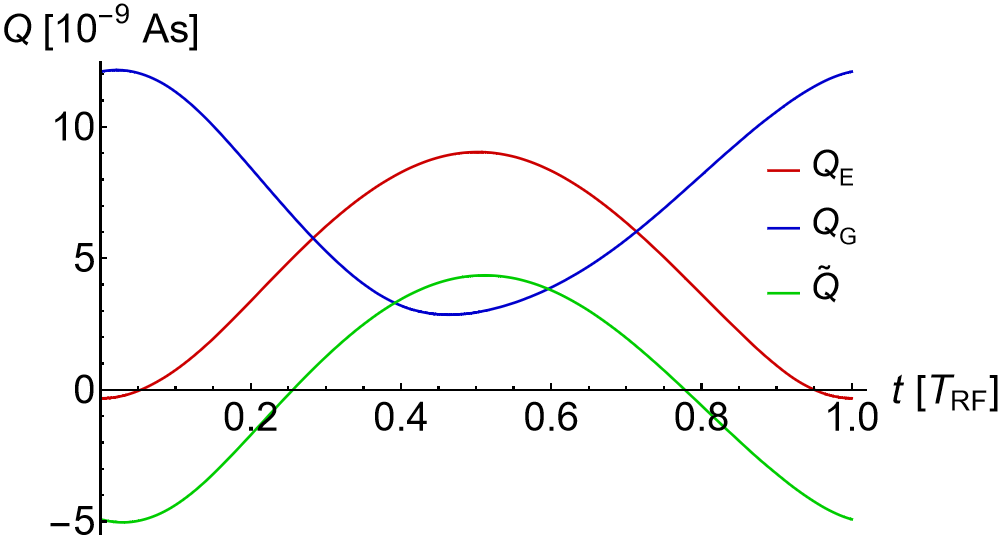}
\else
\fi
	\caption{Charges of the electrode sheath $Q_\mathrm{E}(t)$ and the ground 
	sheath $Q_\mathrm{G}(t)$ together with the average-free integral $\tilde{Q}(t)$
	of the current $I(t)$ for the magnetized  case. \FINAL}
	\label{ChargesMagnetizedCase}
\end{figure}
\begin{figure}[h!]\centering
\iffigures	
\includegraphics[width=15cm]{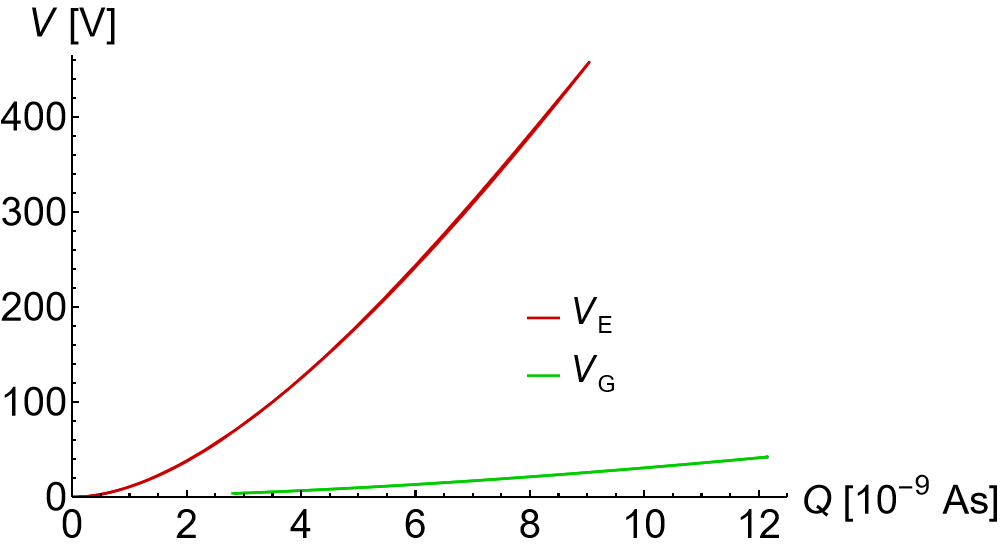}
\else
\fi
\caption{Charge-voltage curves $V_\mathrm{E}(Q)$ of the electrode sheath and $V_\mathrm{G}(Q)$ of the grounded sheath of the magnetized case. \FINAL}
	\label{VQMagnetized}
\end{figure}

Fig.~\ref{VoltagesMagnetizedCaseNEW} shows the voltages.
At first glance, not much has changed compared to the reference.\LB
The applied voltage $V_\mathrm{RF}(t)$ is the same. The discharge voltage $V_\mathrm{d}(t)$, the sheath voltages $V_\mathrm{E}(t)$ and $V_\mathrm{G}(t)$,  
and the self bias 
$V_\mathrm{sb}(t)$ are similar. 
The bulk voltage $V_\mathrm{b}(t)$ remains small.\LB
Obviously, the discharge is capacitive as well, and the quasi-static analysis
\eqref{VoltageBalance} still applies.
In fact, it is now even more applicable: Because of the  higher value of $\omega_\mathrm{PSR}$, the PSR is no \LB
longer excited by the sheath nonlinearities, 
and the total current $I(t)=2\pi H r j(r,t)$ is better captured by the quasi-static
model or the Drude model, Fig.~\ref{CurrentMagnetizedCase}.
As shown in Figs.~\ref{WeightedParticleCurrentMagnetizedCase}
and \ref{WeightedDisplacementCurrentMagnetizedCase},\LB
it is carried by the electrons in the bulk and 
flows as displacement current in the sheaths.
(Due to the absence of PSR oscillations, there is almost no displacement current in the bulk.)\LB
In contrast to the unmagnetized case, there is now also an 
azimuthal current $j_\theta$, see Fig.~\ref{jthetaCurrent}.
\LB
The waveforms of the sheath charges are similar to those of the unmagnetized case, Fig.~\ref{ChargesMagnetizedCase}.
They are, however, larger by a factor of 2.
This ratio, which reflects the higher plasma density\LB
and the thinner sheaths of the magnetized case,
is also visible in the amplitude of the current and in the form 
of the charge voltage relations $V_\mathrm{E}(Q)$
and  $V_\mathrm{G}(Q)$ shown in Fig.~\ref{VQMagnetized}.

The simulation
determines the current-voltage phase angle to $\cos\varphi= \cos 83.5\text{°} = 0.113$
and the  dissipated RF power to $\bar{P} =  6.76\,\mathrm{W}$.
(The Drude model, 
with  $\cos\varphi = \cos 88.8\text{°} = 0.021$ and $\bar{P} =  1.30\,\mathrm{W}$ is far off.)
$\bar{P}_\mathrm{i}=3.98\,\mathrm{W}$ is taken by the ions, 
$\bar{P}_\mathrm{e}=2.78\,\mathrm{W}$ by the electrons. \LB
The power input ratio matches the density ratio, 
$n_\mathrm{max}\mathrm{(M)}/n_\mathrm{max}\mathrm{(U)} \approx \bar{P}_\mathrm{e}\mathrm{(M)}/\bar{P}_\mathrm{e}\mathrm{(U)}
 \approx
8$. \LB
The plasma density of the magnetized case is
higher because the heating  
is more efficient. 
We will demonstrate that it also has a very different physical character.

\begin{figure}[h!]\centering
\iffigures	
   \includegraphics[width=15cm]{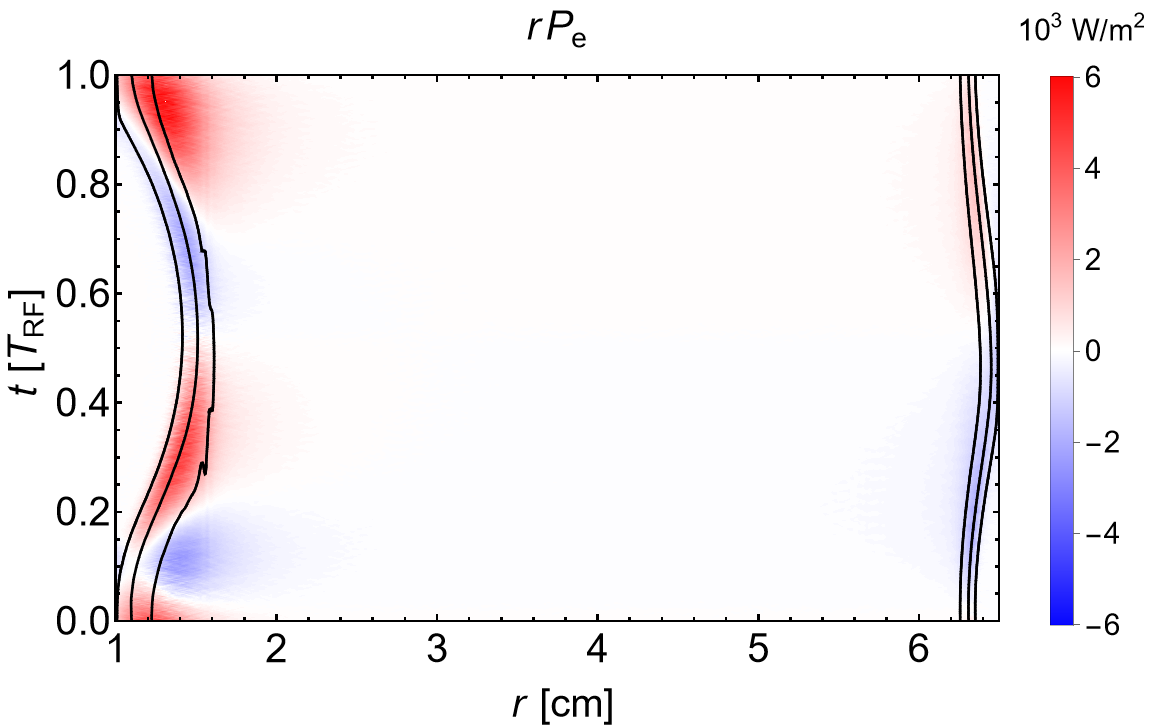} 
\else
\fi
	\caption{The radially weighted spatio-temporally resolved electron power absorption. \FINAL}
	\label{PowerDissipationMagnetizedCase}
\end{figure}

As in the unmagnetized reference case, the heating mechanism is related to the electric field in the ambipolar zone which is neglected in the quasi-static analysis and which is only poorly represented by the Drude model, see Fig.~\ref{ElectricFieldMagneticCase}.
Its individual components can be analyzed by solving the radial momentum balance \eqref{rMomentumBalance} with respect to the electric field:
\begin{align}\label{Eanalysis}
 \! \!\!\!\!  E_r &= -\frac{1}{e n}\!\left( \frac{\partial}{\partial t}(m n u_r) 
    +\frac{1}{r} \frac{\partial}{\partial r}(r m n u_r^2)-
\frac{m n u_\theta^2}{r}\right) 
    - \frac{1}{r e n}\!\left( 
\frac{\partial}{\partial r}\left(r p_{rr} \right)- p_{\theta\theta}\!\right)
- u_\theta B_z
    + \frac{F_r}{e n} \nonumber \\[0.5ex]
    & \equiv E_{\mathrm{i}}+E_{\mathrm{p}}
    +E_{\mathrm{H}} + E_{\mathrm{c}}.
\end{align}
The terms on the right can be obtained from the simulation. Two contributions dominate, the pressure field $E_{\mathrm{p}}$ and the Hall field $E_{\mathrm{H}}$.
The pressure field appears at the electron edges of both  sheaths and points outward, i.e., is negative around $s_\mathrm{E}(t)$ and positive around $s_\mathrm{G}(t)$. \LB
As in the unmagnetized case, it acts as an indicator of the beginning depletion field with which the electrons are roughly in Boltzmann equilibrium.
The Hall field $E_{\mathrm{H}}$ reflects the Lorentz force and appears only in the magnetized zone.
It has a 180° phase shift with respect\LB
to the electron edge $s_\mathrm{E}$. This can be explained
in terms of the single-particle picture above: 
\LB 
As \eqref{Deltarp} shows,
a positive (reversed) field is needed to move electrons to smaller values of $r$ \LB
and a negative one to push them in the opposite direction.)
There is thus a positive interference between $E_\mathrm{p}$ and $E_\mathrm{H}$ 
at the time of the sheath maximum and a cancellation around time of the the sheath minimum.
However, as $E_{\mathrm{H}}$ is also active where quasineutrality holds,  a strong positive
field appears during the sheath minimum in the zone $1\,{\rm cm} \leq r \leq 2\,{\rm cm}$.
The other electric field terms on the right of \eqref{Eanalysis} -- reflecting the influence of electron inertia and collisions -- are much smaller 
(see \cite{WangWenHartmannDonkoDerzsiWangSongWangSchulze2020}). 
The dominance of the pressure field and the Hall field can also be detected
in Fig.~\ref{PowerDissipationMagnetizedCase} which displays a phase-resolved 
map of the radially\LB weighted electron power input
$rP_\mathrm{e} = r j_{\mathrm{e}r} E_r$. 
Its individual constituents, defined on the 
basis of the electrical field analysis
\eqref{Eanalysis}, are displayed in Fig.~\ref{Fig_pow_abs_magn}:
\begin{align}
\!\!\!\!\! r P_r   &= mu_r\!\left( r   \frac{\partial}{\partial t}(n u_r)+ 
\frac{\partial}{\partial r}\left(r n u_r^2 \right) - n u_\theta^2\right)+
u_r \left(
\frac{\partial}{\partial r}\left(r p_{rr} \right)  - 
  p_{\theta\theta} \right)
- r j_r u_\theta B_z
    - r u_r\dot{\pi}_r  \nonumber \\
&\equiv r P_{\mathrm{i}r} +r P_{\mathrm{p}r} 
   + r P_{\mathrm{H}} + r P_{\mathrm{c}r} \label{PowerMagnetizedCase}
\end{align}
The pressure term $P_{\mathrm{p}r}$ appears 
only in the thin  
transition zones around the electron edges;
\LB its value is positive when the sheaths are expanding and negative when they are
retracting.\LB The phase average is close to zero. The Hall term $P_{\mathrm{H}}$ is only active in the magnetized region, \LB
it is negative just after the sheath collapse and positive just before.
The other contributions to the power balance are essentially negligible. 

\begin{figure}[h!]\centering
\iffigures	
   \includegraphics[width=\textwidth]{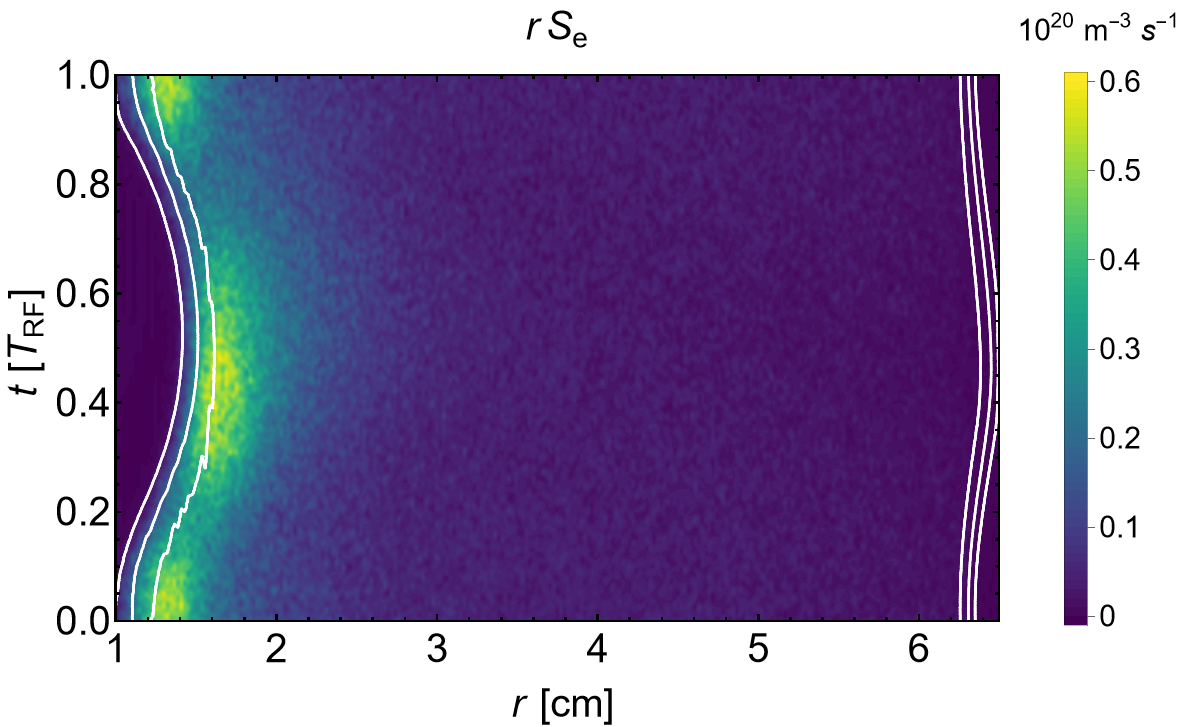} 
\else
\fi
	\caption{Radially weighted ionization rate $r S_\mathrm{e}$ for the magnetized discharge. \FINAL}
	\label{Fig_pow_abs_magn_av}
\end{figure}

The Hall contribution $P_\mathrm{H}$ does not correspond to true power dissipation, as magnetic fields do not do any work and cannot contribute to heating. 
Instead, the term represents a flow of mechanical energy conducted from radial to azimuthal degrees of freedom and back.\LB 
This view is supported by the following equation which is obtained by multiplying the azimuthal momentum balance \eqref{thetaMomentumBalance} with $u_\theta$:
\begin{align}
 r e n u_{\theta} u_{r} B_z   &= u_{\theta}\!\left( r  \frac{\partial}{\partial t}(m n u_{\theta}) %
+ \frac{1}{r}\frac{\partial}{\partial r}\left(r^2 m n u_{r}u_{\theta}\right)\right)+u_{\theta} \frac{1}{r}\frac{\partial}{\partial r}\left(r^2 p_{r\theta}\right)
- r u_{\theta}\dot{\pi}_\theta \\
\nonumber
&= r P_{\mathrm{i}\theta} + r P_{\mathrm{p}\theta}
   + r P_{\mathrm{c}\theta}
\end{align}
The term on the left is the negative of $P_\mathrm{H}$; 
it acts as power source for the azimuthal motion. \LB
Fig.~\ref{Fig_pow_abs_magn} shows the three contributions on the right. 
To begin with the least important one: \LB
The term $P_{\mathrm{p}\theta}$ reflects the shear component of the pressure tensor 
and is linked to the \LB collisionless effect of gyroviscosity: 
Radial transport of the axial momentum component caused by electron gyration 
\cite{hazeltine_2004}. Physically, and by numerical value, it is rather insignificant. 
\LB
The term $P_{\mathrm{i}\theta}$ is associated with electron inertia 
and reflects the periodic acceleration and deceleration of the electrons 
related to the $\vec{E}\!\times\!\vec{B}$ drift. 
During the field growth before the sheath collapse,
energy is transferred to the electrons; after the collapse, 
is transferred back. 
This energizing mechanism is collisionless; it was already captured in the 
single-particle view. \LB
It generates a strong electron beam in the azimuthal direction,
with a velocity that may exceed the thermal velocity significantly. (Even if the drift speed $v_{E\times B}$ is not larger than the thermal velocity but comparable to it, the kinetic energy can be up to four times larger than the thermal energy alone when the gyro motion and the drift have the same direction.)\LB
A sizable number of electrons close to the powered electrode 
can therefore reach energies in the inelastic range,
so that they can participate directly
 in excitation or 
ionization processes. 
\LB
This is seen in the term $P_{\mathrm{c}\theta}$ 
and in the ionisation rate shown  in Fig.~\ref{Fig_pow_abs_magn_av}.
The confining effect of the magnetic field prevents the electrons from being ejected from the magnetized zone,
\LB allowing them to experience large electric fields over a long period of time. 
Note that the number of energetic electrons
in such a zone is increased by the production of new electrons via the ionization process. Such electrons experience the same strong electric field and thus gain the same large $\mathbf{E}\!\times\! \mathbf{B}$ azimuthal drift velocity resulting in a large total kinetic energy. Under a sufficiently large ionization probability during a time interval when an electron feels a large electric field close to the powered electrode, this might significantly enhance the ionization efficiency by creating an ionization avalanche.

\begin{figure}[h!]\centering
\iffigures	
   \includegraphics[width=\textwidth]{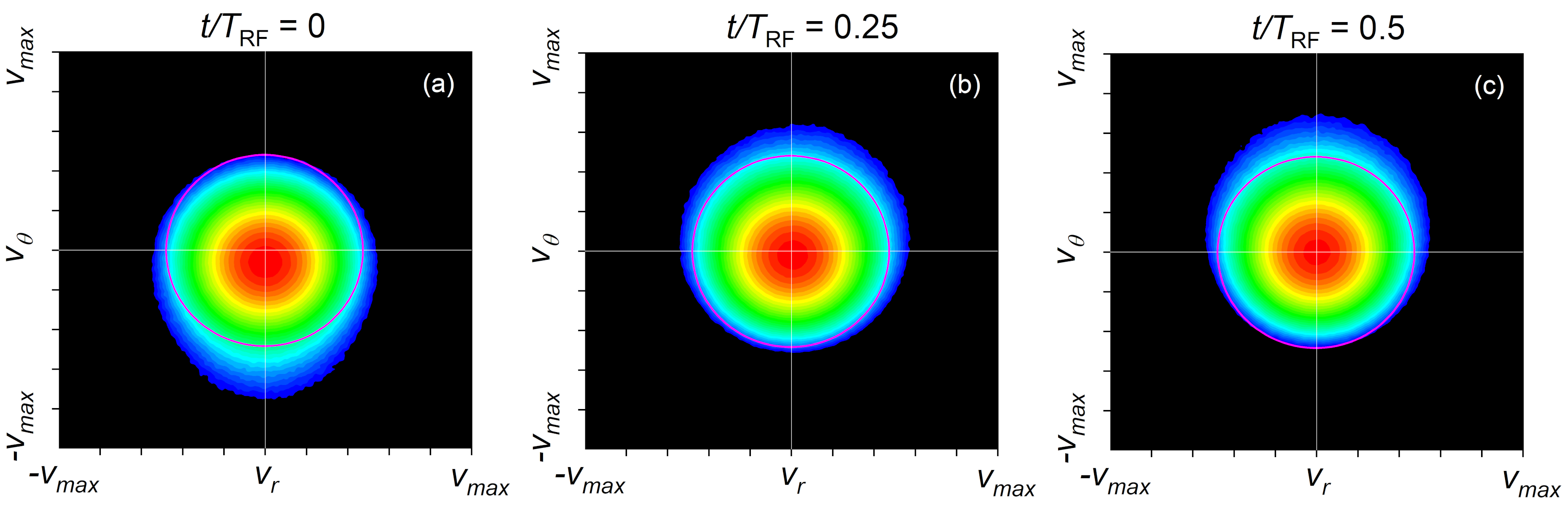} 
\else
\fi
	\caption{Distribution of electrons in the $(v_r,v_\theta)$ plane of the velocity space sampled at different phases of the sheath
	evolution. The magenta circle indicates boundary of the inelastic energy range and corresponds to the $11.5$ eV argon excitation level. \FINAL}
	\label{Fig_EVDF_snapshots}
\end{figure}

The findings are supported by the graphs of Fig.~\ref{Fig_EVDF_snapshots}, which illustrate the electron velocity distribution in the $(v_r,v_\theta)$ plane of the velocity space for different phases of the RF cycle.
 \LB
The electrons were sampled from a segment close to the powered electrode, $1\,{\rm cm} \leq r \leq 2\,{\rm cm}$, where most of the energetic electrons are produced. (Secondary electrons are not shown.) \LB
The velocity range is the interval $[-v_\mathrm{max},v_\mathrm{max}]$,
$v_\mathrm{max}=4.2\times 10^6\,{\rm m/s}$ corresponds to $50\,{\rm eV}$. \LB
The boundary between the elastic and inelastic energy range is shown as a magenta circle.
At $t=0$,  all electrons experience a strong positive (``reversed'') electric field which generates a fast $\vec{E}\!\times\!\vec{B}$ drift in the negative $\theta$ direction (Fig.~\ref{Fig_EVDF_snapshots}a). A substantial fraction of electrons is pushed beyond the inelastic energy threshold. 
When the powered electrode sheath expands \LB
-- we chose the moment $t=0.25\,T_\mathrm{RF}$   --,
the electric field at the sheath edge changes polarity and the $\vec{E}\!\times\!\vec{B}$ drift becomes small for the electrons close to the expanding sheath edge. \LB The corresponding distribution of energetic electrons becomes symmetric in $v_\theta$ (Fig.~\ref{Fig_EVDF_snapshots}b).
\LB
Finally, at $t=0.5\,T_\mathrm{RF}$, when the powered sheath is maximal,
 electrons at the sheath edge experience a strong negative electric field.
 The fast $\vec{E}\!\times\!\vec{B}$ drift in the positive $\theta$  
 direction generates a 
 large number of energetic electrons. 
Electrons closer to the bulk experience only \LB a weak electric field, so their distribution, which dominates the low-energy part, is symmetric.
The slight asymmetry of the energetic tail with respect to $v_r$ 
(Fig.~\ref{Fig_EVDF_snapshots}b)
   can be attributed to magnetized stochastic heating \cite{LiebermanLichtenbergSavas1991}.
   The effects is much weaker than Hall heating.

\begin{figure}[h!]\centering
\iffigures	
   \includegraphics[width=\textwidth]{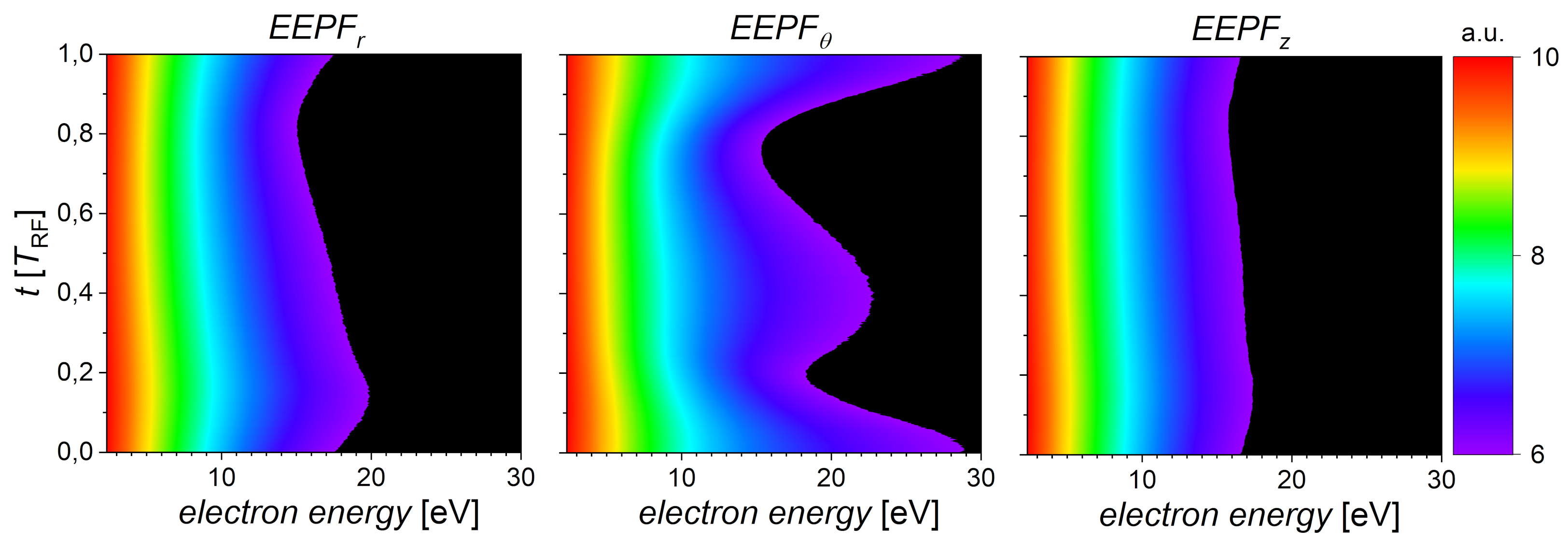} 
\else
\fi
	\caption{Phase-resolved electron distribution in different directions for the magnetized case.\LB Secondary electrons are not displayed.  \FINAL}
	\label{Fig_EEPF_vs_time}
\end{figure}

The described mechanisms are also seen in  
Fig.~\ref{Fig_EEPF_vs_time} which shows the  phase-resolved electron energy probability functions (EEPFs) in the three  directions extracted from the simulation.
\LB (In each case, the full EEPF was integrated  over the 
other directions.) 
The azimuthal $\mathrm{EEPF}_\theta$\LB shows the most pronounced modulation.
Its energetic phases are at $t=0$ and $t=T_\mathrm{RF}/2$, \LB
clearly related to the maxima of the $\vec{E}\!\times\!\vec{B}$-drift 
during sheath collapse and sheath expansion.
The radial $\mathrm{EEPF}_r$ is less modulated; the magnetized sheath heating  (as proposed in \cite{LiebermanLichtenbergSavas1991}) 
and the isotropizing effect of collisions are relatively weak.
The axial $\mathrm{EEPF}_z$, finally,  stays nearly constant,
it is only influenced by the collisions.

\pagebreak

\section{Summary and conclusions \label{s4}}

In this study, we investigated the electron dynamics and the mechanisms of power absorption \LB 
in a radio-frequency-driven, magnetically-enhanced capacitively-coupled plasma (MECCP).
The device in focus was a
cylindrical magnetron with a radially nonuniform magnetic field in axial direction and an electric field in radial direction.
The applied voltage was $V_\mathrm{RF} = 300\,\mathrm{V}$,
\LB
the gas argon  at a pressure of $p=0.5\,\mathrm{Pa}$.
An unmagnetized discharge of the same geometry\LB
and operation conditions was used for comparison.
The dynamics was studied analytically with the cold plasma model and 
a single-particle formalism, and numerically with the inhouse
energy and charge conserving PIC/MCC code ECCOPIC1S-M.

The reference discharge showed the well-known mechanisms of pressure heating, NERH,
and,  to a lesser extent, Ohmic heating,
all acting mainly in the vicinity of the powered sheath.
\LB
The magnetized CCP, in contrast, operates by means of a
significantly more efficient power absorption mechanism, 
which we named ''Hall heating". 
It is  caused by the discharge's need to ensure the electron 
current continuity against the inhibitory effect of the magnetic field. \LB
A single-particle study emphasized the role of polarization drift in cross-field transport.
\LB
%
%
The required strong electric field has a phase shift of 180° compared to the electron edge $s_\mathrm{E}$;\LB
it is negative when the sheath
width 
is maximal and positive (reversed) when it is minimal. 
The corresponding strong azimuthal $\mathbf{E}\!\times\! \mathbf{B}$-drift 
then constitutes the "Hall heating" as such. 
It is by a factor $\omega_\mathrm{ce}/\omega_\mathrm{RF}$ larger than the
polarisation drift and can accelerate a relatively large number of electrons  
into the inelastic energy range.
These electrons -- which are drawn not only from the vicinity of the 
electron edge but also from 
momentarily quasineutral regions -- \LB can either directly participate in inelastic processes such as impact ionization or convert their kinetic energy into
random motion through elastic collisions. 
Hall heating is different from the heating mechanism proposed by Lieberman, Lichtenberg, and Savas
\cite{LiebermanLichtenbergSavas1991},
as it does not rely on multiple collisions with the expanding sheath.
It also differs from Ohmic heating; since it is not a matter of a diffusion process in  energy space.
It is something entirely new.   
We propose to call it the ``$\mu$-mode'', to separate it 
from other heating modes in gas discharges. 
Contribution of this mechanism to the production of energetic \LB electrons participating in the ionization processes can be significantly enhanced due to the fact that the electrons created in the ionization will be energized by the same mechanism via gaining a strong azimuthal $\mathbf{E}\!\times\! \mathbf{B}$ drift, which may result in a ionization avalanche.\LB
A companion study \cite{eremin_2021} will investigate 
the newly described mechanism in a more realistic planar magnetron geometry
and focus also on the electron motion along the magnetic field. \LB
A second companion study \cite{berger_2021} will present experimental observations obtained in a planar RF magnetron and interpret them with help of the newly gained knowledge.

\pagebreak

\color{black}

\section{Acknowledgments}

The authors gratefully acknowledge support by DFG (German Research Foundation) within the framework of the Sonderforschungsbereich SFB-TR 87 and the project
"Plasmabasierte Prozessführung von reaktiven Sputterprozessen" No.~417888799.


\newpage

\begin{appendix}

\section{Approximate solution of the equations of motion}

The electron motion in the cross-sectional $r$-$\theta$ plane
under the influence of a static magnetic field $\vec{B}=B_z(r)\vec{e}_z$ and a dynamic electric field $\vec{E}= E_r(r,t)\vec{e}_r$ can be   
viewed as an inter\-play between gyrorotation and drift.
For a dimensionless notation, we employ the length scale $L$ \LB
of the system and the time scale $T$ of the applied  RF modulation and write the fields and the particle coordinates as
\begin{align}
	 &E_r(r,t) =  \frac{1}{\eta} \frac{m_\mathrm{e} L}{e\, T^{2}}\tilde{E}\! \left(r/L,t/T\right),\\[0.50ex]
   &B_z(r) = \frac{1}{\eta} \frac{m_\mathrm{e}}{e\,T} \tilde{B}\! \left(r/L\right),\\[0.50ex]
	&r(t) = L\tilde{r}(t/T),\\[0.50ex]
	&\theta(t) = \tilde{\theta}(t/T),\\[0.50ex]
	&v_r(t) = \frac{L}{T}\tilde{v}_r(t/T),\\[0.50ex]
	&v_\theta(t) = \frac{L}{T}\tilde{v}_\theta(t/T).
\end{align}
The quantity $\eta \approx 0.1$ (in the magnetized region) is a dimensionless smallness parameter. The adopted scaling 
causes the gyromotion to be fast compared to the RF frequency, 
\begin{align}
  \omega_\mathrm{ce} = \frac{e B_z}{m_\mathrm{e}} =    \frac{1}{\eta} \frac{1}{T} \tilde{B}
     \sim \frac{1}{\eta T},
\end{align}
and the $E\!\times\! B$ drift comparable to the thermal speed which is of order $L/T$,
\begin{align}
    v_{E\times B} = \frac{E_r}{B_z} =  \frac{\tilde{E}}{\tilde{B}} \frac{L}{T} \sim \frac{L}{T}.  
\end{align}
Switching to dimensionless space and time coordinates, $r\to L \tilde{r}$ and $t\to T \tilde{t}$, and then dropping the tilde, 
we formulate the equations of motion as
\begin{align}
&\frac{\mathrm{d}r}{\mathrm{d}t} = v_r,\\[0.50ex]
&\frac{\mathrm{d} v_r}{\mathrm{d}t} = -\frac{1}{\eta}\bigl(E(r,t)+ B(r) v_\theta\bigr) +\frac{v_\theta^2}{r},\\[0.50ex]
&\frac{\mathrm{d}\theta}{\mathrm{d}t}  = \frac{v_\theta}{r},\\[0.50ex]
&\frac{\mathrm{d} v_\theta}{\mathrm{d}t}= \frac{1}{\eta}  B(r)v_r -\frac{v_\theta v_r}{r}.
\end{align}

We introduce the flux function $\Psi(r)$ which is related to the vector potential $A_\theta$,
\begin{equation}
	 \Psi(r) = \int_0^r B(r^\prime)\, r^\prime \mathrm{d}r^\prime
	   = r  A_\theta(r),
\end{equation}
and express the magnetic field as
\begin{equation}
B(r) =  \frac{1}{r} \frac{\partial\Psi}{\partial r}.
\end{equation}
Owing to the  cylindrical symmetry of the configuration, there is an exact constant of motion, 
the canonical momentum $p_\theta$ in $\theta$ direction. 
It is used to define a reference radius $\hat{r}$: 
\begin{equation}
	p_\theta = r v_\theta  - \frac{1}{\eta}\Psi(r)   \stackrel{\displaystyle !}{=}  - \frac{1}{\eta}\Psi(\hat{r}). \label{CanonicalMomentum}
\end{equation}
The electric and magnetic fields at the reference radius $\hat{r}$ are also
denoted by a circumflex; the spatial derivatives with an 
additional prime:
\begin{align}
    &\hat{E}(t) = E\left(\hat{r},t\right), \\
    &\hat{B} = B\left(\hat{r}\right), \\[0.25ex]
    &\hat{E}^\prime(t) = \frac{\partial E}{\partial r}\left(\hat{r},t\right), \\[0.25ex]
    &\hat{B}^\prime =\frac{\partial B}{\partial r}\left(\hat{r}\right), 
\end{align}

Equation \eqref{CanonicalMomentum} can be used to eliminate the variable $v_\theta$ via \begin{align}
    v_\theta = \frac{\Psi(r)-\Psi(\hat{r})}{\eta\hspace{0.5mm} r}.
    \label{thetaElimination}
\end{align}
The remaining equations are then
\begin{align}
      &\frac{\mathrm{d}r}{\mathrm{d}t} = v_r,\\[0.25ex]
      &\frac{\mathrm{d}\theta}{\mathrm{d}t} = \frac{\Psi(r)-\Psi(\hat{r})}{\eta\hspace{0.5mm} r^2},\\[0.25ex]
      &\frac{\mathrm{d}v_r}{\mathrm{d}t} = -\frac{E(r,t)} {\eta}
           +\frac{(\Psi(r)-\Psi(\hat{r}))^{2}}{\eta^2 r^{3}}-\frac{(\Psi(r)-\Psi(\hat{r})) \Psi^{\prime}(r)}{\eta^2 r^{2}}.
\end{align}
To account for the time scale disparity, we distinguish between the fast gyroscale
$t_\mathrm{f} = \eta t$ and the slower RF scale $t_\mathrm{s} = t$. Splitting the time derivative
accordingly, we obtain:
\begin{align}
      &\frac{1}{\eta}\frac{\mathrm{\partial}r}{\mathrm{\partial}t_\mathrm{f}} +\frac{\mathrm{\partial}r}{\mathrm{\partial}t_\mathrm{s}} = v_r,\\[0.25ex]
      &\frac{1}{\eta}\frac{\mathrm{\partial}\theta}{\mathrm{\partial}t_\mathrm{f}} +\frac{\mathrm{\partial}\theta}{\mathrm{\partial}t_\mathrm{s}} = \frac{\Psi(r)-\Psi(\hat{r})}{\eta\hspace{0.5mm} r^2},\\[0.25ex]
      &\frac{1}{\eta}\frac{\mathrm{\partial}v_r}{\mathrm{\partial}t_\mathrm{f}} +\frac{\mathrm{\partial}v_r}{\mathrm{\partial}t_\mathrm{s}} = -\frac{E(r,t_\mathrm{s})} {\eta}
           +\frac{(\Psi(r)-\Psi(\hat{r}))^{2}}{\eta^2 r^{3}}-\frac{(\Psi(r)-\Psi(\hat{r})) \Psi^{\prime}(r)}{\eta^2 r^{2}}.
\end{align}
We now make a power series ansatz in $\eta$, 
                   displaying only those terms that are actually used. 
Note that the leading order of $r$ is fixed to be the reference radius $\hat{r}$:
\begin{align}
    & r(t_\mathrm{f},t_\mathrm{s}) = \hat{r} + \eta\, r^{(1)}(t_\mathrm{f},t_\mathrm{s})
                                   + \eta^2\, r^{(2)}(t_\mathrm{f},t_\mathrm{s})+\ldots,\\
    & \theta(t_\mathrm{f},t_\mathrm{s}) = \theta^{(0)}(t_\mathrm{f},t_\mathrm{s}) + \eta\,  \theta^{(1)}(t_\mathrm{f},t_\mathrm{s}) 
    + \ldots, \\
    & v_r(t_\mathrm{f},t_\mathrm{s}) = v_r^{(0)}(t_\mathrm{f},t_\mathrm{s})
                                   + \eta\, v_r^{(1)}(t_\mathrm{f},t_\mathrm{s}) + \ldots.
\end{align}

Expanding the equations of motion into a Taylor series in the smallness parameter $\eta$ and sorting for powers gives a hierarchy of equations. In leading order, they are:
\begin{align}
   &\frac{\partial r^{(1)}}{\partial t_\mathrm{f}} - v_r^{(0)} = 0, \\[0.25ex]
   &\frac{\partial \theta^{(0)}}{\partial t_\mathrm{f}} = 0, \\[0.25ex]
   &\frac{\partial v_r^{(0)}}{\partial t_\mathrm{f}} + \hat{B}^2 r^{(1)} =  - \hat{E}(t_\mathrm{s}).
\end{align}
This system can readily be solved. The integration constants $\rho$, $\hat\phi$,
and $\hat\theta$ may still depend on 
the slow time $t_\mathrm{s}$ (which we suppress in the notation for brevity):
\begin{align}
  &  r^{(1)}(t_\mathrm{f},t_\mathrm{s}) = \rho\cos\bigl(\hat{B} t_\mathrm{f}+\hat\phi\bigr) -\frac{\hat{E}}{\hat{B}^2},\\[0.25ex]
  &  \theta^{(0)}(t_\mathrm{f},t_\mathrm{s}) = \hat{\theta}, \\[0.25ex] 
  &  v_r^{(0)}(t_\mathrm{f},t_\mathrm{s}) = - \hat{B} \rho\cos\bigl(\hat{B} t_\mathrm{f}+\hat\phi\bigr).
\end{align}
The dynamical equations of the next order appear as
inhomogeneous differential equations for the 
quantities $r^{(2)}$, $\theta^{(1)}$, and $v_r^{(1)}$. Note that the homogeneous
part is formally identical to the equations of the leading order:
\begin{align}
&\frac{\partial r^{(2)}}{\partial t_\mathrm{f}} - v_r^{(1)} 
=
\frac{1}{\hat{B}^2} \frac{\partial\hat{E}}{\partial t_\mathrm{s}}        
-\cos\bigl(\hat{B} t_\mathrm{f}+\hat{\phi}\bigr) \frac{\partial\rho}{\partial t_\mathrm{s}}+\rho \sin\bigl(\hat{B} t_\mathrm{f}+\hat{\phi}\bigr) \frac{\partial\phi}{\partial t_\mathrm{s}}, \\[0.25ex]
 &  \frac{\partial \theta^{(1)}}{\partial t_\mathrm{f}}
       = - \frac{\partial \hat\theta}{\partial t_\mathrm{s}} 
       + \frac{\hat{B} }{\hat{r}}
       \rho(t_\mathrm{s})\cos\bigl(\hat{B} t_\mathrm{f}+\hat{\phi}\bigr)- \frac{\hat{E}}{\hat{r}\hat{B}}, \\[0.25ex]
 & \frac{\partial v_r^{(1)}}{\partial t_\mathrm{f}} + \hat{B}^2 r^{(2)}=    \frac{\hat{E} \hat{E}^\prime}{\hat{B}^{2}}+\frac{3}{2}\left(\frac{1}{\hat{r}}-\frac{\hat{B}^\prime}{\hat{B}}\right)\frac{\hat{E}^2}{\hat{B}^2}
 +\left(3\left(\frac{\hat{B}^\prime}{B}-\frac{1}{\hat{r}}\right)\hat{E}-\hat{E}^\prime\right) \rho \cos\bigl(\hat{B} t_\mathrm{f}+\hat{\phi}\bigr)
 \\ \nonumber 
 & \hspace{3cm}
 +\left(\frac{3 \hat{B}^2}{4 \hat{r}}-\frac{3 \hat{B}\hat{B}^\prime}{4}\right) \rho^{2}+\left(\frac{3\hat{B}^2}{4\hat{r}}-\frac{3 \hat{B}\hat{B}^\prime}{4}\right) \rho^{2}
 \cos\left(2\bigl(\hat{B} t_\mathrm{f}+\hat{\phi}\bigr)\right)
 \\\nonumber
  & \hspace{3cm}+ \hat{B}\left( \sin\bigl(\hat{B} t_\mathrm{f}+\hat{\phi}\bigr) \frac{\partial\rho}{\partial t_\mathrm{s}} +  
\rho\cos\bigl(\hat{B} t_\mathrm{f}+\hat{\phi}\bigr)  \frac{\partial\hat{\phi}}{\partial t_\mathrm{s}}\right).
\end{align}

When solving these equations, care must be taken to avoid terms that 
linearly diverge in $t_\mathrm{f}$.\LB 
This poses consistency conditions on the inhomogeneous 
terms on the right, which can be solved for the evolution 
equations for the integration constants $\rho$, $\hat{\phi}$, and $\hat{\theta}$ in the time $t_\mathrm{s}$.
It~turns out that $\rho$ is a constant altogether. 
(This fact is related to the adiabatic constancy of the magnetic moment $\mu = v_\perp^2/2B$ which is valid for drift theories under 
general conditions.)
The angles $\hat{\phi}$, and $\hat{\theta}_0$, in contrast, exhibit slow drifts:
\begin{align}
& \frac{\partial\rho}{\partial t_\mathrm{s}} = 0,\\[0.25ex]
&    \frac{\partial\hat{\phi}}{\partial t_\mathrm{s}} = 
\frac{3}{2}
\left(
\frac{1}{ \hat{B} \hat{r}} -
\frac{  \hat{B}^\prime  }{ \hat{B}^2 }\right)\hat{E} +
 \frac{1}{2 \hat{B} }\hat{E}^\prime, \\[0.25ex]
  & \frac{\partial\hat\theta}{\partial t_\mathrm{s}} = -\frac{\hat{E}}{\hat{B}\hat{r}}.
\end{align}
Under this condition, the next order quantities can be calculated as
\begin{align}
 r^{(2)}(t_\mathrm{f},t_\mathrm{s}) =&
 \frac{\hat{E}\hat{E}^\prime} {\hat{B}^{4}}
 -\frac{3 \hat{B}^\prime \hat{E}^2}{2 \hat{B}^5} +\frac{3 \hat{E}^2}{2 \hat{B}^4 \hat{r}}
  + \left(\frac{3 \hat{B}^\prime \hat{E}}{4 \hat{B}^3}-\frac{\hat{E}^\prime}{4 \hat{B}^2}-\frac{3 \hat{E}}{4 \hat{B}^{2} \hat{r}}\right) \rho\cos\bigl(\hat{B} t_\mathrm{f}+\hat{\phi}\bigr)\\[0.25ex]\nonumber
  &
 +\left(-\frac{3 \hat{B}^\prime}{4 \hat{B}}+\frac{3}{4 \hat{r}}+
\left(\frac{\hat{B}^\prime}{4\hat{B}}-\frac{1}{4\hat{r}}\right)\rho^2 
\cos\left(2 \bigl(\hat{B} t_\mathrm{f}+\hat{\phi}\bigr)\right)\right) \\ \nonumber
   & +\rho_a \cos\bigl(\hat{B} t_\mathrm{f}+\phi\bigr)+\rho_b \sin\bigl(\hat{B} t_\mathrm{f}+\hat{\phi}\bigr), \\[0.25ex]
   \theta^{(1)}(t_\mathrm{f},t_\mathrm{s}) =& \frac{1}{\hat{r}} \rho\sin\bigl(\hat{B} t_\mathrm{f}+\hat{\phi}\bigr) + \hat{\theta}_1,\\[0.25ex]
    v_r^{(1)}(t_\mathrm{f},t_\mathrm{s}) =&
 -\frac{1}{\hat{B}^{2}} 
 \frac{\partial \hat{E}}{\partial t_\mathrm{s}} + 
 \left(\frac{3 \hat{B}^\prime\hat{E}}{4 \hat{B}^2}-\frac{\hat{E}^\prime}{4 \hat{B}}-\frac{3\hat{E}}{4 \hat{B}\hat{r}}\right)\rho\sin\bigl(\hat{B} t_\mathrm{f}+\hat{\phi}\bigr) \\\nonumber &
 +\left(\frac{\hat{B}}{2\bar{r}}-\frac{\hat{B}^\prime}{2}\right)\rho^2\sin\left(2 \bigl(\hat{B} t_\mathrm{f}+\hat{\phi}\bigr)\right)\\[0.25ex]\nonumber
 &
 + \hat{B}\left(\rho_b\cos\bigl(\hat{B} t_\mathrm{f}+\hat{\phi}\bigr)-\rho_a \sin\bigl(\hat{B} t_\mathrm{f}+\hat{\phi}\bigr)\right).
\end{align}
Using rule \eqref{thetaElimination}, the first two orders of the
azimuth velocity $v_\theta$ can be reconstructed as
\begin{align}
    v_\theta^{(0)}(t_\mathrm{f},t_\mathrm{s}) 
    =& -\frac{\hat{E}}{\hat{B}}+\hat{B} \rho \cos\bigl(\hat{B} t_\mathrm{f}+\hat{\phi}\bigr), \\[0.25ex]
 v_\theta^{(1)}(t_\mathrm{f},t_\mathrm{s}) =&
 \frac{\hat{E}\hat{E}^\prime}{\hat{B}^3} 
    -\frac{\hat{B}^\prime \hat{E}^2}{\hat{B}^4}+\frac{\hat{E}^2}{\hat{B}^3 \hat{r}}+\left(-\frac{\hat{B}^\prime \hat{E}}{4 \hat{B}^2}-\frac{\hat{E}^\prime}{4\hat{B}}+\frac{\hat{E}}{4 \hat{B} \hat{r}}\right) \rho \cos\bigl(\hat{B} t_\mathrm{f}+\hat{\phi}\bigr)\\[0.25ex]\nonumber & +\rho^{2}\left(-\frac{\hat{B}^\prime}{2}+\frac{\hat{B}}{2 \hat{r}}+\left(\frac{\hat{B}^\prime}{2}-\frac{\hat{B}}{2 \hat{r}}\right) \cos\left(2 \bigl(\hat{B} t_\mathrm{f}+\hat{\phi}\bigr)\right)\right) \\[0.25ex]\nonumber
    & +  \hat{B}\left(\rho_a  \cos\bigl(\hat{B} t_\mathrm{f}+\hat{\phi}\bigr) +
                 \rho_b \sin\bigl(\hat{B} t_\mathrm{f}+\hat{\phi}\bigr)\right).
\end{align}

In order to construct the most economical representation possible 
for the electron trajectory, we proceed as follows. First, we note
that the leading order of the expansion can absorb the homogeneous
part of the higher order by redefinition of the integration 
constants $\rho$, $\hat{\phi}$, and $\hat{\theta}$.\LB
Second, we restrict ourselves to contributions of first order in either gyromotion or drift. \LB
This leads to the form: 
\begin{align}
    & r(t_\mathrm{f},t_\mathrm{s}) = \hat{r} + \eta\,\biggl( \rho\cos\bigl(\hat{B} t_\mathrm{f}+\hat\phi\bigr) -\frac{\hat{E}}{\hat{B}^2}\biggr),\\
    & \theta(t_\mathrm{f},t_\mathrm{s}) = \hat{\theta} +  \eta\,\frac{1}{\hat{r}} \rho\sin\bigl(\hat{B} t_\mathrm{f}+\hat{\phi}\bigr) , \\
    & v_r(t_\mathrm{f},t_\mathrm{s}) = - \hat{B} \rho\cos\bigl(\hat{B} t_\mathrm{f}+\hat\phi\bigr) -\eta \frac{1}{\hat{B}^{2}} 
 \frac{\partial \hat{E}}{\partial t_\mathrm{s}},\\
    & v_\theta(t_\mathrm{f},t_\mathrm{s}) = -\frac{\hat{E}}{\hat{B}}+\hat{B} \rho \cos\bigl(\hat{B} t_\mathrm{f}+\hat{\phi}\bigr). 
    \end{align}
For the integration constants, we have the slow drift
\begin{align}
 &    \frac{\partial\hat\phi}{\partial t_\mathrm{s}} = 
\frac{3}{2}
\left(
\frac{1}{ B \hat{r}} -
\frac{  B^\prime  }{ B^2 }\right)\hat{E} +
 \frac{1}{2 B }\hat{E}^\prime, \\[0.5ex]
  & \frac{\partial\hat\theta}{\partial t_\mathrm{s}} = -\frac{\hat{E}}{\hat{B}\hat{r}}.
\end{align}
Lastly, we remove the two-timescale formalism and the normalization to obtain the desired approximation of the electron trajectories: 
\begin{align}
    & r(t) = \hat{r}  - \frac{m_\mathrm{e}\hat{E}}{e\hat{B}_r^2}+  \rho\cos\biggl( \frac{e \hat{B}}{m_\mathrm{e}}t +\hat\phi\biggr),\\
    & \theta(t) = \hat{\theta} +  \frac{\rho}{\hat{r}} \sin\biggl( \frac{e \hat{B}}{m_\mathrm{e}}t +\hat\phi\biggr), \\
    & v_r(t) =- \frac{m_\mathrm{e}}{e\hat{B}^{2}} 
 \frac{\partial \hat{E}}{\partial t_\mathrm{s}} -  \frac{e \hat{B}}{m_\mathrm{e}} \rho\cos\biggl( \frac{e \hat{B}}{m_\mathrm{e}}t +\hat\phi\biggr) ,\\
    & v_\theta(t) =
    -\frac{\hat{E}}{\hat{B}}+ \frac{e \hat{B}}{m_\mathrm{e}} \rho\cos\biggl( \frac{e \hat{B}}{m_\mathrm{e}}t +\hat\phi\biggr).
    \end{align}
The parameters $\hat{r}$ and $\rho$ are constants; the parameters $\hat{\phi}$
and $\hat{\theta}$ follow drift equations:
\begin{align}
&    \frac{\partial\hat{\phi}}{\partial t} = 
\frac{3}{2}
\left(
\frac{1}{\hat{B} \hat{r}} -
\frac{  \hat{B}^\prime  }{ \hat{B}^2 }\right)\hat{E} +
 \frac{1}{2 \hat{B} }\hat{E}^\prime, \\[0.25ex]
  & \frac{\partial\hat\theta}{\partial t}=-\frac{\hat{E}}{\hat{B} \hat{r}}.
\end{align}
By introducing the reference gyrofrequency $\hat{\Omega} = e\hat{B}/m_\mathrm{e}$, the drift velocity $\hat{v}_{E\times B} = \hat{E}/\hat{B}$, and its spatial derivative $\hat{v}^\prime_{E\times B} = \hat{E}/\hat{B}^\prime$,
 the representation of a trajectory can be written
\begin{align}
    & r(t) = \hat{r}  - \frac{\hat{v}_{E\times B}}{\hat{\Omega}}+  \rho\cos\bigl(\hat{\Omega}t +\hat\phi\bigr),\\
    & \theta(t) = \hat{\theta} +  \frac{\rho}{\hat{r}} \sin\bigl(\hat{\Omega}t +\hat\phi\bigr), \\
    & v_r(t) = - \frac{1}{\hat{\Omega}} 
 \frac{\mathrm{d}\hat{v}_{E\times B}}{\mathrm{d} t}-  \hat{\Omega} \rho\cos\bigl(\hat{\Omega}t +\hat\phi\bigr) ,\\
    & v_\theta(t) =
    -\hat{v}_{E\times B}+ \hat{\Omega} \rho\cos\bigl(\hat{\Omega} +\hat\phi\bigr),
    \end{align}
and the drift equations for the integration constants read:
\begin{align}
&    \frac{\partial\hat{\phi}}{\partial t} = 
\frac{3}{2}
\left(
\frac{1}{\hat{r}} -
\frac{  \hat{B}^\prime  }{ \hat{B} }\right)\hat{v}_{E\times B} +
 \frac{1}{2}\hat{v}^\prime_{E\times B}, \\[0.25ex]
  & \frac{\partial\hat\theta}{\partial t}=-\frac{\hat{v}_{E\times B}}{\hat{r}}.
\end{align}


\end{appendix}

\pagebreak

\bibliographystyle{ieeetr}

\bibliography{references}

\pagebreak

\pagebreak

\clearpage
\pagebreak

\clearpage
\pagebreak

\clearpage
\pagebreak

\pagebreak

\pagebreak

\clearpage
\pagebreak

\pagebreak

\clearpage
\pagebreak

\clearpage
\pagebreak

\clearpage
\pagebreak

\pagebreak

\pagebreak

\clearpage
\pagebreak

\clearpage

\pagebreak

\clearpage
\pagebreak

\clearpage
\pagebreak

\clearpage
\pagebreak

\clearpage
\pagebreak

\vfill

\clearpage

\pagebreak

\clearpage


\pagebreak

\clearpage
\pagebreak

\clearpage
\pagebreak

\pagebreak

\clearpage
\pagebreak

\clearpage
\pagebreak

\clearpage
\pagebreak

\clearpage
\pagebreak

\clearpage
\pagebreak

\clearpage
\pagebreak

\clearpage
\pagebreak

\clearpage
\pagebreak

\clearpage
\pagebreak



%
%


\clearpage
\pagebreak


\pagebreak

\begin{figure}[h!]\centering
\iffigures	
   \includegraphics[width=\textwidth]{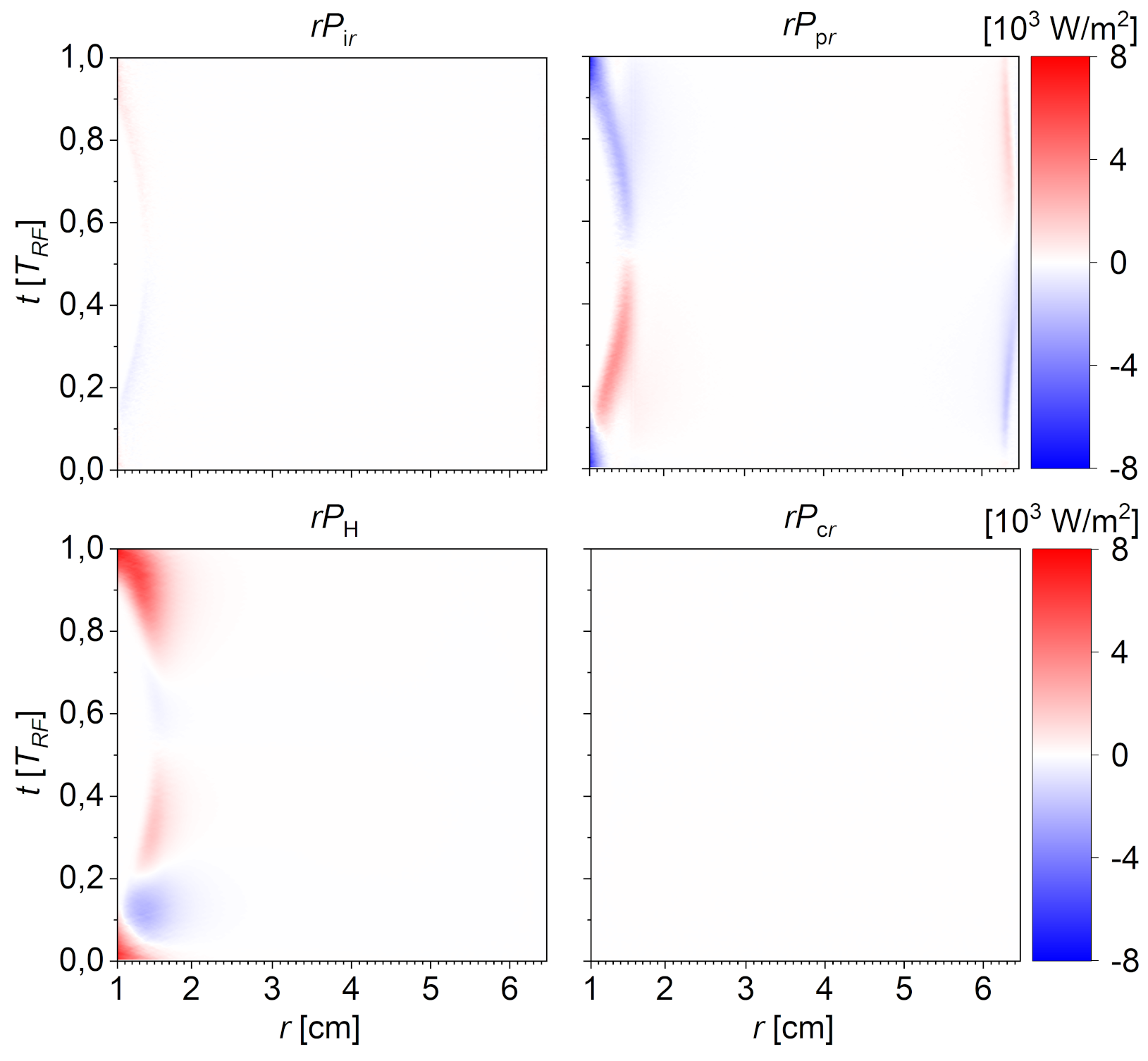} 
\else
\fi
	\caption{The radially weighted spatio-temporally resolved constituents from \eqref{PowerMagnetizedCase} of the power dissipation. \FINAL
}
	\label{Fig_pow_abs_magn}
\end{figure} 

\pagebreak

\begin{figure}[h!]\centering
\iffigures	
   \includegraphics[width=\textwidth]{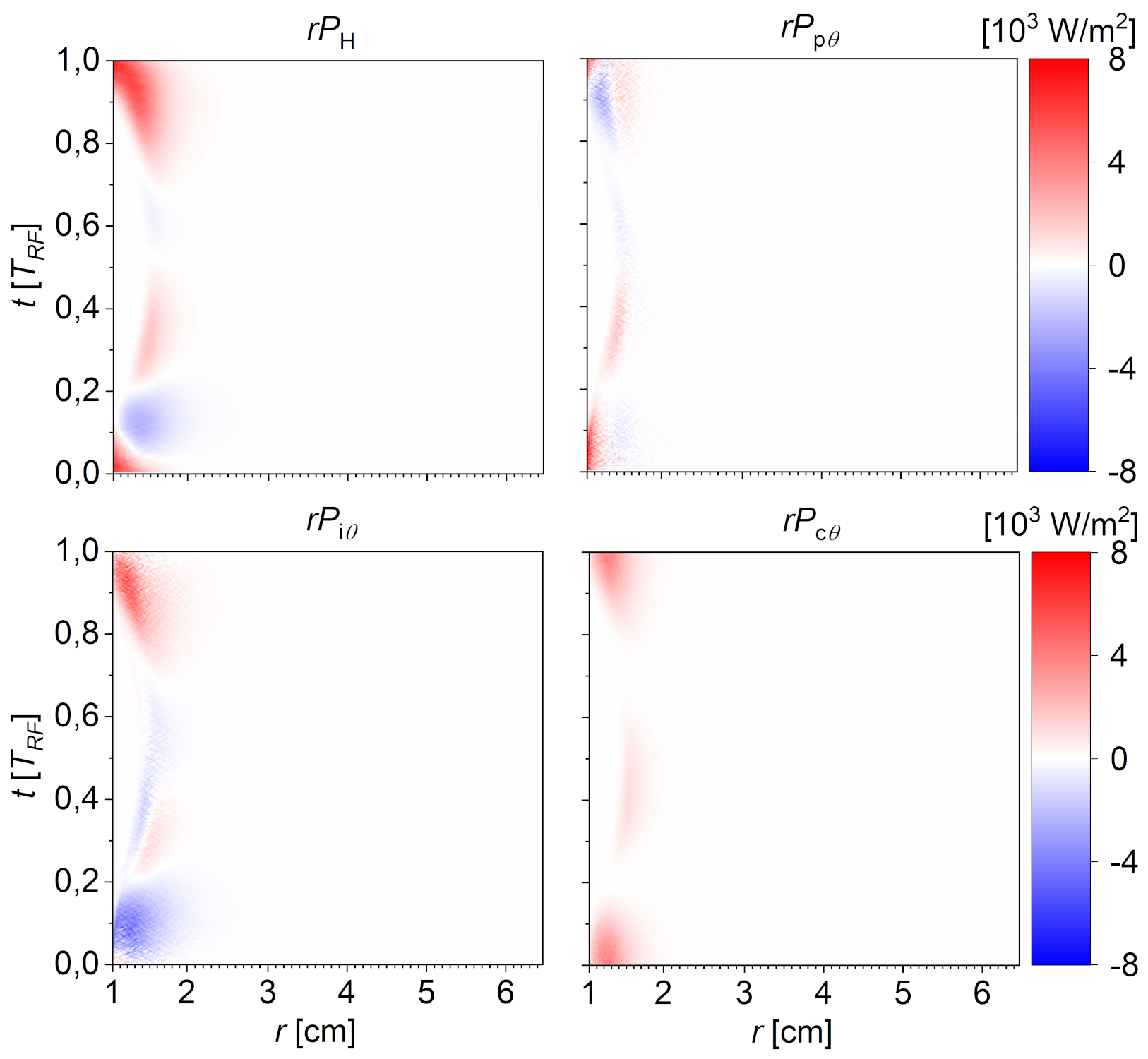} 
\else
\fi
	\caption{The spatio-temporally resolved Hall power and its constitutents. }
	\label{Fig_pow_abs_magn}
\end{figure} 

\pagebreak

\pagebreak

\pagebreak

%


\end{document}